\documentclass[twocolumn,showpacs,preprintnumbers,amsmath,amssymb]{revtex4}
\usepackage{epsfig}
\usepackage{amsmath}

\begin{document}

\title{Inward and Outward Node Accessibility in Complex Networks as 
Revealed by Non-Linear Dynamics}

\author{Luciano da Fontoura Costa}
\affiliation{Institute of Physics at S\~ao Carlos, University of
S\~ao Paulo, PO Box 369, S\~ao Carlos, S\~ao Paulo, 13560-970 Brazil}

\date{9th Jan 2008}

\begin{abstract}
In this work, the outward and inward accessibilities of individual
nodes are defined and their potential for application is illustrated
with respect to the investigation of 6 different types of networks.
The outward accessibility quantifies the potential of an individual
node for accessing other nodes through random walks.  The inward
accessibility measures the potential of a given node of being accessed
by other nodes.  Both the inward and outward accessibilities are
measured with respect to successive time steps along the walks,
providing an interesting means for the characterization of the
transient non-linear dynamics of accessibility.  Self-avoiding walks
are considered here because they are more purposive and necessarily
finite (unlike traditional random walks).  The results include the
identification of the fact that the inward values tend to be much
smaller than the outward values, the tendency of the inward
accessibility to be highly correlated with the node degree while the
outward values are mostly uncorrelated with that measurements, the
distinct behavior of the accessibility in geographical networks, the
dominance of hubs in scale free networks, as well as the enhanced
uniformity of the accessibilities for the path-regular model.  Also
investigated was the possibility to predict the accessibility of a
given node in terms of its respective degree. The concepts of inward
and outward accessibility, as well as the several obtained results,
have several implications and potential for applications to several
real-time problems including disease spreading, WWW surfing, protein
interaction, cortical networks and network attacks, among others.
\end{abstract}

\maketitle

\vspace{0.5cm}
\emph{`A map of Esmeralda should include, marked in different
colored inks, all these routes, solid and liquid, evident and hidden.'
(Invisible Cities, I. Calvino)}

\section{Introduction} 

One of the main current issues in complex systems, defined by an
intersection between dynamic systems and complex networks, regards the
interaction between structure and dynamics (e.g.~\cite{Newman:2003,
Boccaletti:2006}).  A great deal of the current attention has been
concentrated on linear dynamics, especially synchronization
(e.g.~\cite{Boccaletti:2006}), allowing interesting results on how the
several structural features of networks affect this type of dynamics.
Yet, while linear dynamics involving diffusion (related to traditional
random walks) and synchronization are of extreme importance to a wide
range of important natural and artificial systems, while also
providing a first order approximation to several non-linear dynamics,
growing attention is poised to focus the extension of the
structure-dynamics investigations to intrinsically non-linear
dynamical systems.  Previous related approaches include the studies of
associative memory capacity in Hopfield networks with connectivity is
underlain by complex networks~\cite{Stauffer_Hopfield,Stauffer_Costa},
the study of Ising and Potts dynamics in complex networks
(e.g.~\cite{Ising_BA,Ising_BA2,Costa_Ising}), opinion formation
(e.g.~\cite{Sznajd_BA,Ising_BA2,Fortunato_opinion}), among others.

Because of their simplicity of implementation in computer systems,
random walks stand out as a particularly interesting approach to
investigate the relationship between structure and dynamics in
non-linear dynamical systems.  While many works have addressed
traditional random walks in complex networks
(e.g.~\cite{Zhou:2003,Rieger:2004, Masuda:2004, Pons_comm:2005,
Eisler:2005, Costa_corrs:2007}), relatively fewer articles have
reported studies related on non-linear random walks
(e.g.~\cite{Kinouchi_thesaurus:2001,Yang:2005, Costa_know:2006,
Herrero_self:2003, Herrero_self:2005,Costa_know:2006}).  More
recently, the transient evolution of self-avoiding walks on several
types of networks was quantified in terms of diversity entropy
signatures~\cite{Costa_diverse:2008}.  

The current work extends and complements the previous investigations
reported in~\cite{Costa_diverse:2008} by considering the accessibility
of each individual node in the network as quantified by the
frequencies of visits along self-avoiding walks initiating at all the
other nodes and introducing the concept of \emph{inward}
accessibility.  While the concept of diversity adopted in that work
related to the diversity of nodes (not edges) reached by random walks,
here we use the term \emph{accessibility} in order to quantify the
estimated number of accesses received or made by specific nodes along
successive steps of the walks.  Two situations are considered
regarding node accessibility: (i) how many nodes can be potentially
accessed after starting a self-avoiding random walk from a specific
node; and (ii) how many accesses a particular node can potentially
receive from self-avoiding random walks starting from all other nodes.
These two properties are henceforth called
\emph{outward} and \emph{inward accessibilities}.  In both cases, it is 
interesting to quantify objectively the accessibility with respect to
each subsequent number of steps.  The current article proposes
intuitive and sound means to quantify the inward and outward
accessibilities.  The definition of these two features naturally
motivates the investigation of possible asymmetries of accesses,
e.g. quantified by the ratio between the outward and the inward
accessibilities.  We also consider networks from 6 families of models
presenting distinctive structures, namely Erd\H{o}s-R\'enyi,
Barab\'asi-Albert, Watts-Strogatz, a simple geographical type of
network, as well as two knitted networks (PN and PA) introduced
recently~\cite{Costa_comp:2007}.  Several interesting results are
reported and discussed regarding the inward and outward
accessibilities, which are shown to present properties intrinsic to
each type of network.  Another main result is that the outward
accessibility tends to be substantially larger than the inward
accessibility.  In order to try to relate the accessibilities (a
dynamical property of the networks) with structural features, we
consider the possible correlations between the accessibilities and the
degree of each node in the network.  It is shown that though a strong
positive correlation is verified between the inward accessibility and
degree (especially at the later steps of the walks), while the outward
accessibility can only be predicted from the node degree (i.e. strong
correlation between these two measurement) for the very first step of
the self-avoiding walks.

The article starts by presenting the basic concepts, including the
motivation and definition of the inward and outward accessibilities,
and proceeds by presenting and discussing the results with respect to
individual node accessibility and correlations between the latter and
the node degree.

\section{Basic Concepts}

This section presents the basic concepts and methods used in this
article regarding complex network representation and characterization,
complex network models, as well as the definitions of inward and
outward accessibilities and a simple computational method for their
estimation.

\subsection{Network Basics}

Complex networks can be understood as graphs with particularly
intricate structure.  A unweighted, undirected complex network can be
completely represented in terms of its \emph{adjacency matrix} $K$ of
dimension $N \times N$, so that each edge extending from node $i$ to
node $j$ implies $K(i,j)=K(j,i)=1$, with $K(i,j)=K(j,i)=0$ being
otherwise imposed.  The \emph{immediate neighbors} of a node $i$ are
those nodes which are linked to $i$ through a single respective edge.
Two nodes are said to be \emph{adjacent} if one is the immediate
neighbor of the other, and vice versa.  Two edges are said to be
adjacent whenever they have at least one of their extremities attached
to a same node.  The \emph{degree} of a node $i$ corresponds to the
number of its immediate neighbors.  Nodes with degree one are called
\emph{extremity nodes}.

A \emph{walk} is a sequence of adjacent links, starting at an initial
node and proceeding along successive steps $h$.  A \emph{self-avoiding
walk} is a walk which never visits any node or edge more than once.  A
\emph{path} is the subgraph associated to a self-avoiding walk.
Because of their intrinsic nature, paths represent the most economic
way, i.e. involving the smallest number of edges, to interconnect a
set of nodes.  In addition, unlike traditional random walks,
self-avoiding walks necessarily terminate (i.e. the moving agent
always reaches a point from which it can no longer proceed because of
lack of unvisited edges or nodes).  The nodes where the self-avoiding
walks end are henceforth called \emph{termination nodes}. In the
present work, the moving agents are assumed to stay at the terminal
nodes after reaching them.  This choice allows the conservation of the
number of moving agents along all time steps.
See~\cite{Costa_diverse:2008} for an additional discussion of such an
assumption.

\subsection{Complex Network Models}

Six different types of networks~\cite{Albert_Barab:2002,
Dorogov_Mendes:2002, Newman:2003, Boccaletti:2006,
Costa_surv:2007,Costa_comp:2007} are considered in this article,
exhibiting markedly distinct structural properties.  The
Erd\H{o}s-R\'enyi (ER) (see also~\cite{Flory}) are obtained by
undirectionally connecting $N$ initially isolated nodes with a
constant probability $\gamma$.  The average degree of this network is
$\left< k \right> = (N-1)\gamma$.  A Barab\'asi-Albert (BA) network
can be obtained by starting with $m0$ nodes and incorporating
additional nodes, one at a time, with $m$ edges, which are attached to
the previous nodes with probability proportional to their respective
degrees.  The average degree of this network is $\left< k \right> =
2m$.  The Watts-Strogatz networks (WS) can be generated by starting
with a completely regular network, henceforth a linear regular
structure, where each node is connected to its neighbors at each side.
Afterwards, a percentage of edges (in our case 10
of existing edges) are randomly rewired.  The geographical type of
network considered in this article is obtained by distributing $N$
nodes uniformly along a two-dimensional space and connecting every par
of nodes with distance is smaller or equal to a maximum distance
$d_{max}$.  The two knitted networks~\cite{Costa_comp:2007} used in
this article are the path-transformed BA network (PA) and the
path-regular network (PN).  The former is obtained by transforming the
star-connections in a BA network into respective
paths~\cite{Costa_comp:2007,Costa_path:2007}, so that the resulting
network contains a power law distribution of path lengths.  The latter
type, i.e. path-regular network, is particularly simple and is
obtained by applying the following procedure several times: (a) start
with N isolated nodes; (b) choose an initial node and make it the
current node; (c) select randomly, but without repetition, a new node
among the remaining nodes and connect it to the current node; (d)
repeat step (c) until all nodes are connected through a self-avoiding
path.  This kind of network has been found~\cite{Costa_comp:2007,
Costa_longest:2007} to exhibit remarkable uniformly with respect to a
large series of measurements which is much higher than that presented
by the ER model.

All networks are grown so as to have approximately the same average
degree $\left< k \right>$ and number of nodes $N$.  We assume
$K(i,i)=0$ in all networks considered in this work.  Also, only the
largest connected component of each network is considered.  However,
because of the relatively large the average node degree adopted in
this work ($\left< k \right> = 2m = 6$), most nodes will be typically
contained in the principal connected component.

\subsection{Inward and Outward Accessibility}

\begin{figure}[htb]
  \vspace{0.3cm} \begin{center}
  \includegraphics[width=0.9\linewidth]{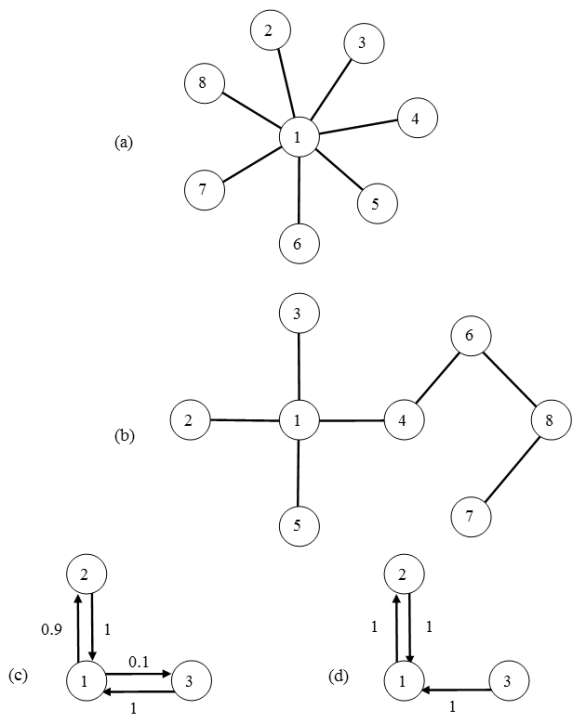}
  \caption{A simple, but fundamental, network (a)involving a central
              node and 7 surrounding nodes.  Other simple networks
              considered for illustration of the accessibilities (b-d).
              Please refer to the text for discussion.}~\label{fig:ex}
  \end{center}
\end{figure}

Consider the simple but particularly important network shown in
Figure~\ref{fig:ex}(a), involving a central node $1$ connected to 7
surrounding nodes, totaling $N=8$ nodes.  Let a moving agent start a
non-preferential self-avoiding walk from node 1.  After one step, the
agent may be at any of the surrounding nodes 2 to 8.  If we repeat
this walk $M$ times and count the number of times $Q(2)$ that the
agent reaches node 2, we can define the rate of accesses to that node
as

\begin{equation}
  r(M) = \frac{Q(2)}{M}
\end{equation}

After one step, the \emph{probability} of reaching node 2 (or any of
the other nodes 3 to 8), henceforth represented as $P_1(1,2)$, is
formally defined as the following limit:

\begin{equation}
  P_1(1,2) = \lim_{M \rightarrow \infty} r(M)  \nonumber 
\end{equation}

More specifically, in the case of Figure~\ref{fig:ex}, the above limit
converges to $P_1(1,2) = 1/7$.  Actually, if we define the sequence of
outcomes of the successive walks after a given number of steps as
corresponding to a visit to node 2 or otherwise, we have a Bernoulli
process with $M$ observations and $p=P_1(1,2)$ to which we can
associate the random variable $Q(2)$ corresponding to number of times
the agent reaches node 2.  It can be shown that the random variable
$Q(2)$ follows the binomial distribution, with mean $\left< Q(2)
\right> = M p$ and standard deviation $\sigma_{Q(2)} =
+\sqrt{Mp(1-p)}$.  We can now define the mean frequency of accesses to
node 2 as

\begin{equation}
  \left< R(2) \right> = \frac{\left< Q(2) \right>}{M} = 
        \frac{M P_1(1,2)}{M} = P_1(1,2) = 1/7  \nonumber 
\end{equation}

which has standard deviation given as $\sigma_{Q(2)}/M$ (observe that
$\lim_{M \rightarrow \infty} \sigma_{Q(2)}/M = 0$).  The mean period
of accesses to node 2 is given as

\begin{equation}
  \left< T(2) \right> = \frac{M}{\left< Q(2) \right>} = 
        \frac{M}{M P_1(1,2)} = \frac{1}{P_1(1,2)} = 7  \nonumber 
\end{equation}

Now, let us consider the situation in which $M$ self-avoiding walks
start not only at node 1, but at \emph{all nodes} (because the walks
do not interact one another, they can be performed simultaneously or
not).  More specifically, $M$ walks are initiated at each node of the
network. We consider that each moving agent avoids only the nodes
already visited by itself, ignoring visits by the other nodes.
Observe that several real systems of interest can be modeled by such
walks, including the disease spreading and WWW surfing.  After one
step, all the moving agents starting from nodes 2 to 8 will
necessarily be at node 1, so that the frequency of access to node 1
remains as before and equal to $1/7$.  A completely different
situation arises at the second step of the walks, in which case we
have:

\begin{eqnarray}
  P_2(1,w) = 0; \nonumber \\ 
  P_2(w,1) = 0;  \nonumber \\
  P_2(v,2) = 1/6;  \nonumber 
\end{eqnarray}

where $w \in \{2, 3, \ldots,8 \}$ and $v \in 3, 4, \ldots,8 \}$.
Because we have a total of $N$ independent self-avoiding walks, it
becomes reasonable to redefine the frequency of accesses to a node
$i$, henceforth called \emph{inward accessibility} of node $i$ after
$h$ steps, as

\begin{equation}
  \left< IA_h(i) \right> = \sum_{q=1, q\neq i}^{N} \frac{
             \left<Q(i) \right>}{M(N-1)} 
             \sum_{q=1,q \neq i}^{N} 
             \frac{\left< P_h(q,i) \right>}{(N-1)} 
\end{equation}

where the normalizing factor $(N-1)$ is used instead of $N$ because
this corresponds to the maximum number of nodes which can reach node
$i$ after a single step (i.e. node $i$ is naturally excluded from the
normalization).  Therefore, the inward accessibility of any node
varies from 0 to 1.  Because of the non-linear dynamics of
self-avoiding walks, the values of $P_h(i,j)$ cannot be easily
calculated in analytical fashion.  In the current article, these
probabilities are calculated computationally as described below.
Observe also that $\sum_{i=1}^{N}P(i,j) = 1$ and, generally, $P_h(i,j)
\ne P_h(j,i)$.

The above definition of the inward accessibility measurement has as
its main motivation the quantification of how accessible a given node
$i$ is by moving agents performing self-avoiding random walks, all of
which starting from all nodes simultaneously.  As the number of such
walks increase (or more agents are liberated simultaneously at each
node), the inward accessibility of a node $i$ will provide an
indication of the intensity of visits to that node.  

Let us now address the motivation and definition of the \emph{outward
accessibility} of a node $i$ after $h$ steps of self-avoiding walks.
Consider again the situation in Figure~\ref{fig:ex}, and let's assume
that a self-avoiding random walk started at node 1.  After the first
step, any of the nodes 2 to 8 will have an equal probability $P_1(1,w)
= 1/7$ of being accessed.  Such a non-preferential, equiprobable
transition from node 1 to the other nodes corresponds to the optimal
situation favoring all the surrounding nodes to be accessed more
uniformly and more speedily as successive (or simultaneous)
self-avoiding walks are performed starting at node 1.  Such an
effectiveness of accesses can be conveniently
expressed~\cite{Costa_diverse:2008} in terms of the entropy
(e.g.~\cite{Latora_entropy}) of the probabilities $P_1(1,w)$.
Therefore, in the above example we have this entropy $E_h(i)$ is given
as:

\begin{equation}
  E_1(1) = - \sum_{q=2}^{N} P_1(1,q) ln(P_1(1,q))  \nonumber 
\end{equation}

Though such a feature provides a sound basis for quantifying the
diversity of the walks originating at $i$, explored partially
in~\cite{Costa_diverse:2008}, it has the shortcoming of exhibiting
values comprehended between 0 and $ln(N-1)$, and not between 0 and 1
as is the case with the above defined inward accessibility.  The
possibility of normalizing $E_h(i)$ by dividing it by $ln(N-1)$
presents the inconvenient that the inward and outward accessibilities
involving transitions with the same probabilities would have different
values.  In order to illustrate such a problem, refer to the situation
depicted in Figure~\ref{fig:ex}(b), where the central node 1 is
connected to 4 surrounding nodes (1, 2, 3 and 4), while other nodes
are connected to node 4.  The inward accessibility and diversity
entropy of node 1 can be calculated as

\begin{eqnarray}
  IA_1(1) = \frac{\sum_{q=2}^{5} P_1(q,1)}{N-1} = 4/7 
                \approx 0.571   \nonumber  \\
  \frac{E_1(1)}{ln(N-1)}   = -ln(0.25)/ln(7) \approx 0.712 \nonumber 
\end{eqnarray}

In order to avoid such a problem, in this work, we therefore define
the \emph{outward accessibility} as

\begin{equation}
  OA_h(i) = \frac{exp(E_h(i))}{N-1}  
\end{equation}

Observe that a node having just one immediate neighbor will have the
smallest entropy value $OA=1/(N-1)$ (we are not considering networks
with a single isolated node).

Now, going back to the example in Figure~\ref{fig:ex}(b), we have:

\begin{equation}
  E_1(1) = -ln(0.25)  \nonumber
\end{equation}

so that 

\begin{eqnarray}
  IA_1(1) = \frac{\sum_{q=2}^{5} P_1(q,1)}{N-1} = 4/7    \nonumber \\
  OA_1(1) = \frac{exp(-ln(0.25))}{7} = 4/7 \nonumber
\end{eqnarray}

i.e. the inward and outward accessibilities will be equal for the
specific situation involving equiprobability transitions from a
central node.  Also, we necessarily have that $0 \leq IA \leq 1$. A
situation involving non-equiprobable transitions is shown in
Figure~\ref{fig:ex}(c).  In this case:

\begin{eqnarray}
  IA_1(1) = \frac{\sum_{q=2}^{3} P_1(q,1)}{N-1} = 2/2 = 1    \nonumber  \\
  OA_1(1) = \frac{1}{exp(0.1 ln(0.1)+0.9 ln(0.9)) (2)} 
                 \approx 0.692 \nonumber
\end{eqnarray}

Such values properly reflect the fact that node 1 in the network in
Figure~\ref{fig:ex}(c) has greater (actually optimal) potential to
receive accesses from other nodes after one step than to access other
nodes after that same number of steps.  Actually, in the limit when
$P_1(1,2)$ becomes very small, we have

\begin{equation}
   \lim_{a \rightarrow 0} \frac{1}{exp(a ln(a)+(1-a)ln(1-a))} = 0.5
\end{equation}

which, as could be expected, corresponds to the situation illustrated
in Figure~\ref{fig:ex}(d).

It is interesting to observe that the quantity $1/exp(-E_h(i))$ can be
understood, from the perspective of outward accessibility, as the
\emph{`equivalent number of nodes'} attached to the reference node
$i$, a real value instead of an integer.  In other words, the outward
accessibility of node 1 in the network in Figure~\ref{fig:ex} would be
equivalent to the accessibility of that node connected to
$1/exp(-E_1(1))$ `fractionary nodes' with equiprobable transitions
$exp(-E_1(1))$.  In that case, we have $1.384$ equivalent nodes and
transition probabilities $0.722$.

It should be observed that the above definitions of the inward and
outward accessibilities are independent of the type of walk under
consideration and can be easy and immediately extended to all other
types of random walks and Markovian systems, linear or not, provided
the transition probabilities $P_h(i,j)$ can be estimated or
calculated.  Actually, such probabilities can be understood as
corresponding to the \emph{strengths} of respectively defined virtual
networks, in the sense that each probability $P_h(i,j)$ can be
interpreted as a virtual link~\cite{Costa_genperc} between nodes $i$
and $j$ at distance $h$.  Consequently, it becomes reasonable to speak
of nodes with high probabilities as \emph{hubs} for the respectie $h$.

Because of the non-linear dynamics of the self-avoiding walks, the
transition probabilities $P_h(i,j)$ between all pairs of distinct
nodes are difficult to be analytically calculated.  In this work, we
adopt the simple computational procedure described
in~\cite{Costa_diverse:2008}, which involves performing several
self-avoiding walks from each node and accumulating the relative
frequencies of access to each visited node in terms of the number of
steps $h$.

\section{Results and Discussion}~\label{sec:results}

A representative network has been chosen for each of the 6 considered
network types in order to illustrate the estimation and interpretation
of the inward and outward accessibilities of nodes.  All networks had
$\left< k \right> \approx 6$ and $N=100$, except for the GG case,
which has $N=91$ nodes.  Because the length of the self-avoiding walks
were observed to be almost always equal to the maximum number of steps
considered in this work (i.e. 10 steps), the obtained results are
reasonably devoid of border effects (i.e. most of the walks did not
reach the extremity nodes).  This does not mean, however, that the
following results will not exhibit specific scaling effects with the
network size $N$.

\begin{figure*}
  \vspace{0.3cm} \begin{center}
  \includegraphics[width=0.95\linewidth]{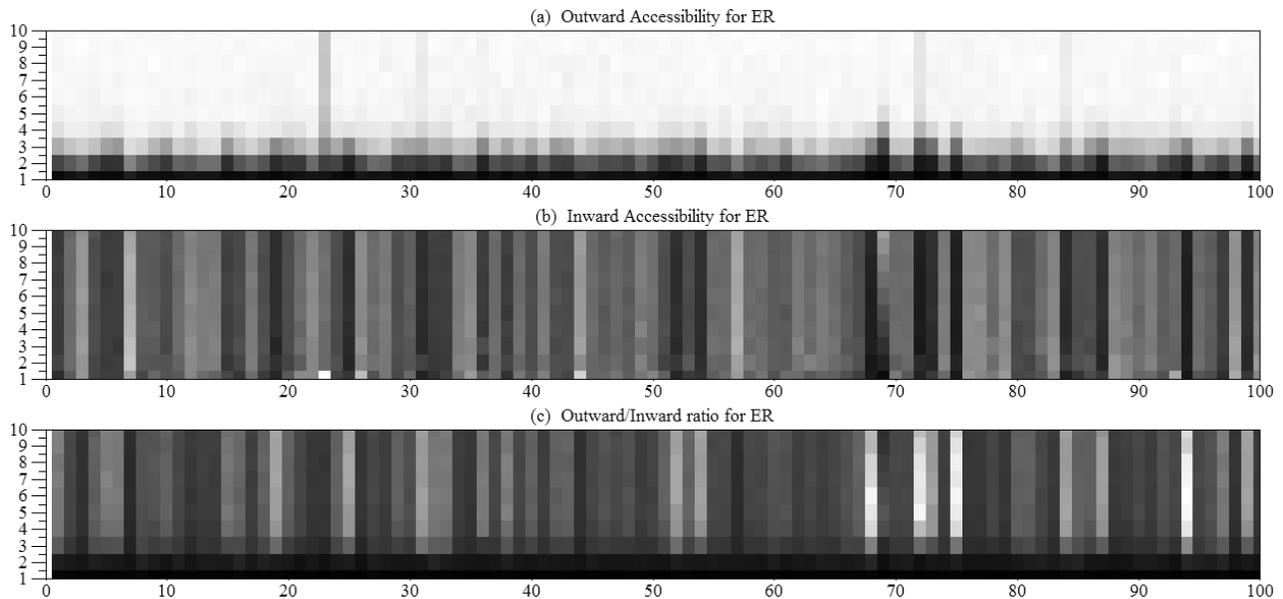} 
  \caption{The outward accessibility (a), inward accessibility (b)
    and the outward/inward ratio (c) for the 10 initial steps 
    obtained for the considered ER network with $N=100$ nodes and
    $\left< k \right> = 6$.}
    ~\label{fig:agram_ER}  \end{center}
\end{figure*}

\begin{figure*}
  \vspace{0.3cm} \begin{center}
  \includegraphics[width=0.95\linewidth]{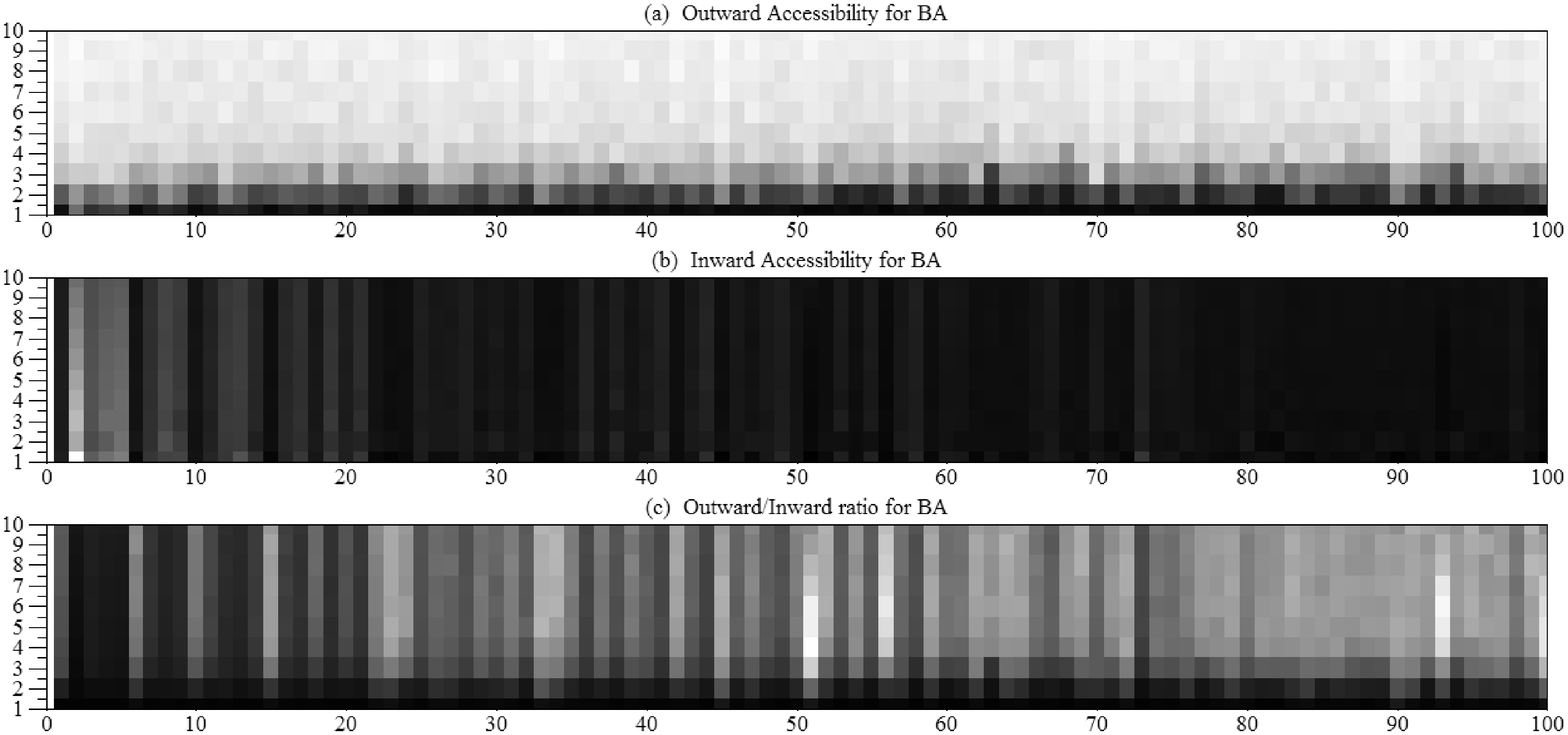}
  \caption{The outward accessibility (a), inward accessibility (b)
    and the outward/inward ratio (c) for the 10 initial steps 
    obtained for the considered BA network with $N=100$ nodes and
    $\left< k \right> = 6$.
    }~\label{fig:agram_BA}  \end{center}
\end{figure*}

\begin{figure*}
  \vspace{0.3cm} \begin{center}
  \includegraphics[width=0.95\linewidth]{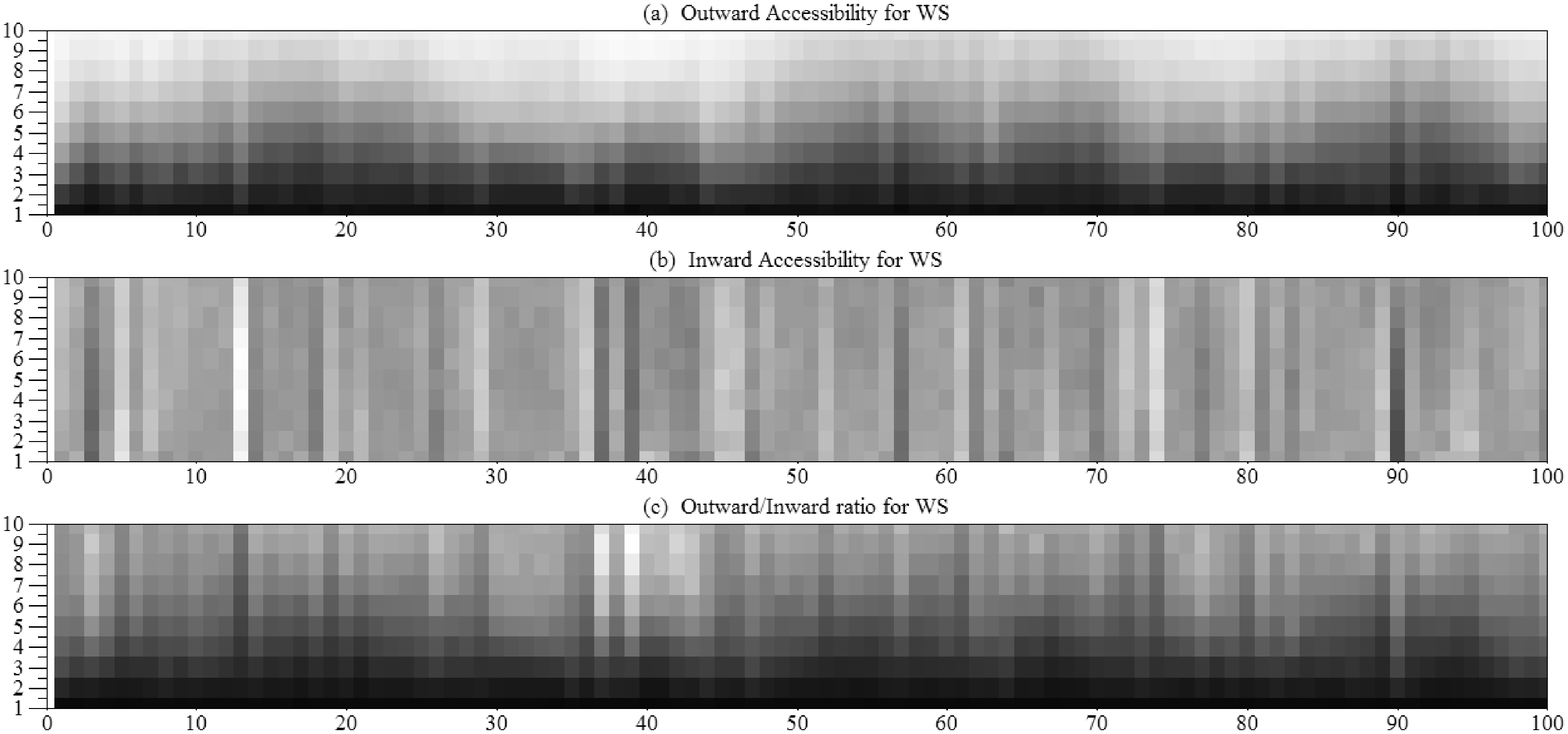}
  \caption{The outward accessibility (a), inward accessibility (b)
    and the outward/inward ratio (c) for the 10 initial steps 
    obtained for the considered WS network with $N=100$ nodes and
    $\left< k \right> = 6$.
    }~\label{fig:agram_WS}  \end{center}
\end{figure*}

\begin{figure*}
  \vspace{0.3cm} \begin{center}
  \includegraphics[width=0.95\linewidth]{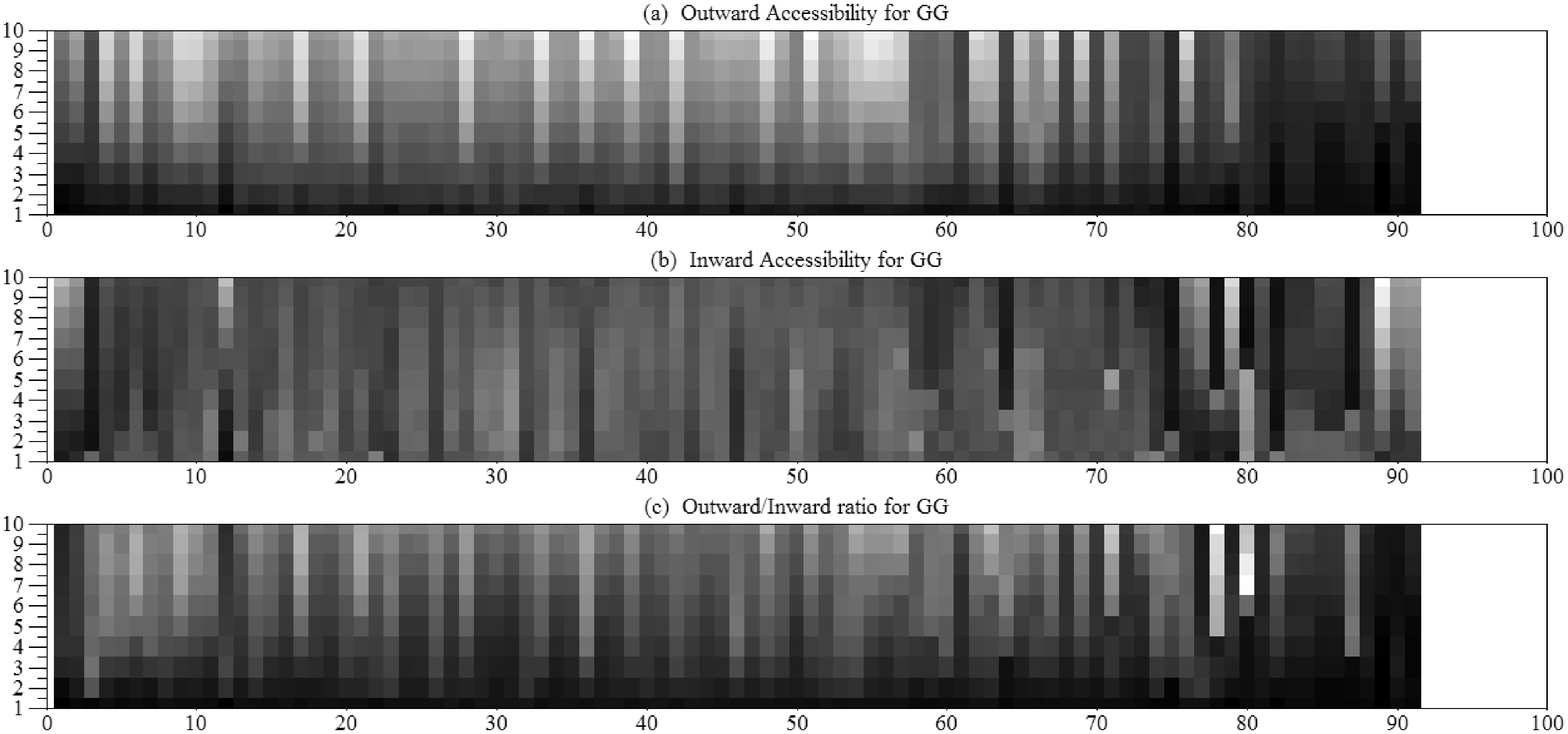}
  \caption{The outward accessibility (a), inward accessibility (b)
    and the outward/inward ratio (c) for the 10 initial steps 
    obtained for the considered GG network with $N=91$ nodes and
    $\left< k \right> = 6$.
    }~\label{fig:agram_GG}  \end{center}
\end{figure*}

\begin{figure*}
  \vspace{0.3cm} \begin{center}
  \includegraphics[width=0.95\linewidth]{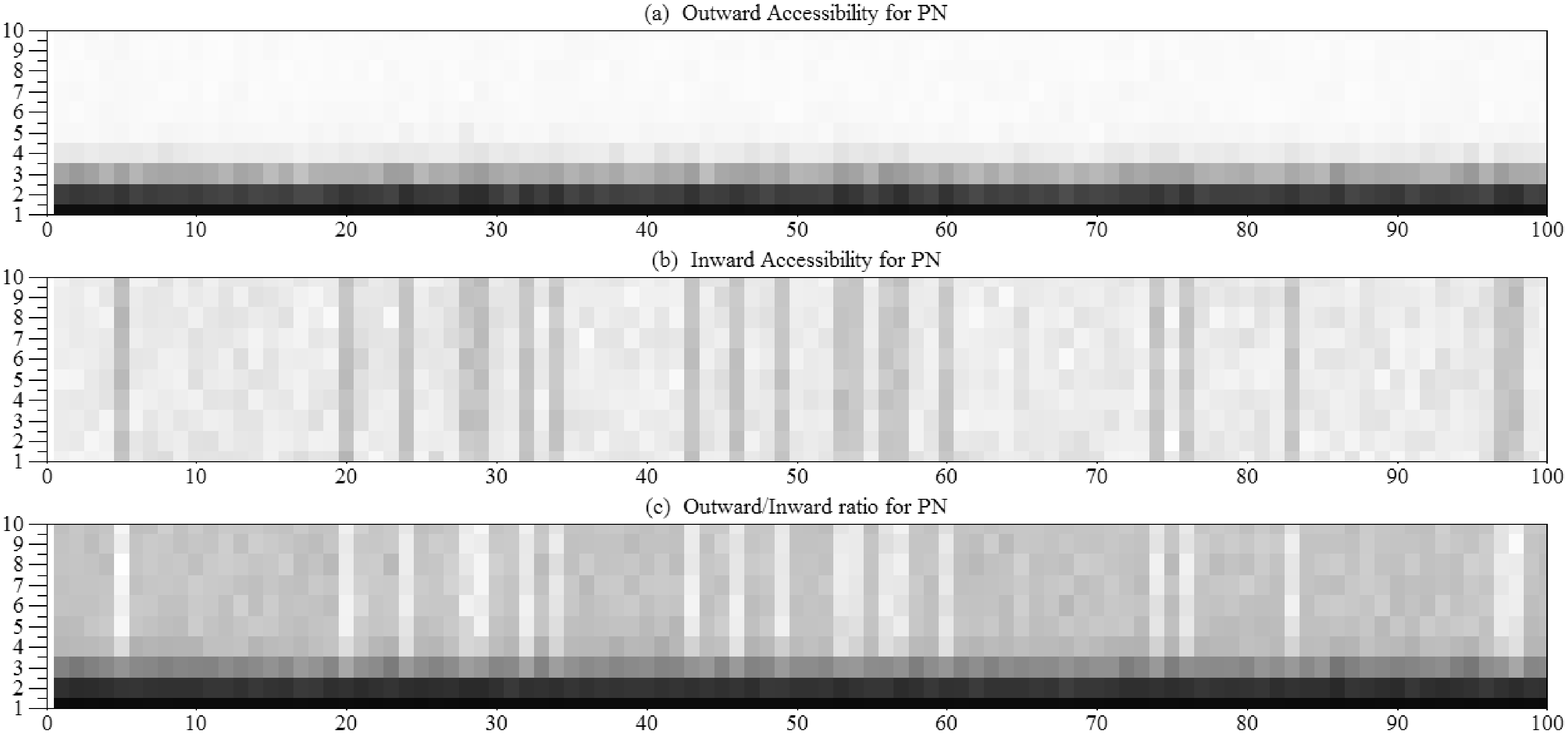}
  \caption{The outward accessibility (a), inward accessibility (b)
    and the outward/inward ratio (c) for the 10 initial steps 
    obtained for the considered PN network with $N=100$ nodes and
    $\left< k \right> = 6$.
    }~\label{fig:agram_PN}  \end{center}
\end{figure*}

\begin{figure*}
  \vspace{0.3cm} \begin{center}
  \includegraphics[width=0.95\linewidth]{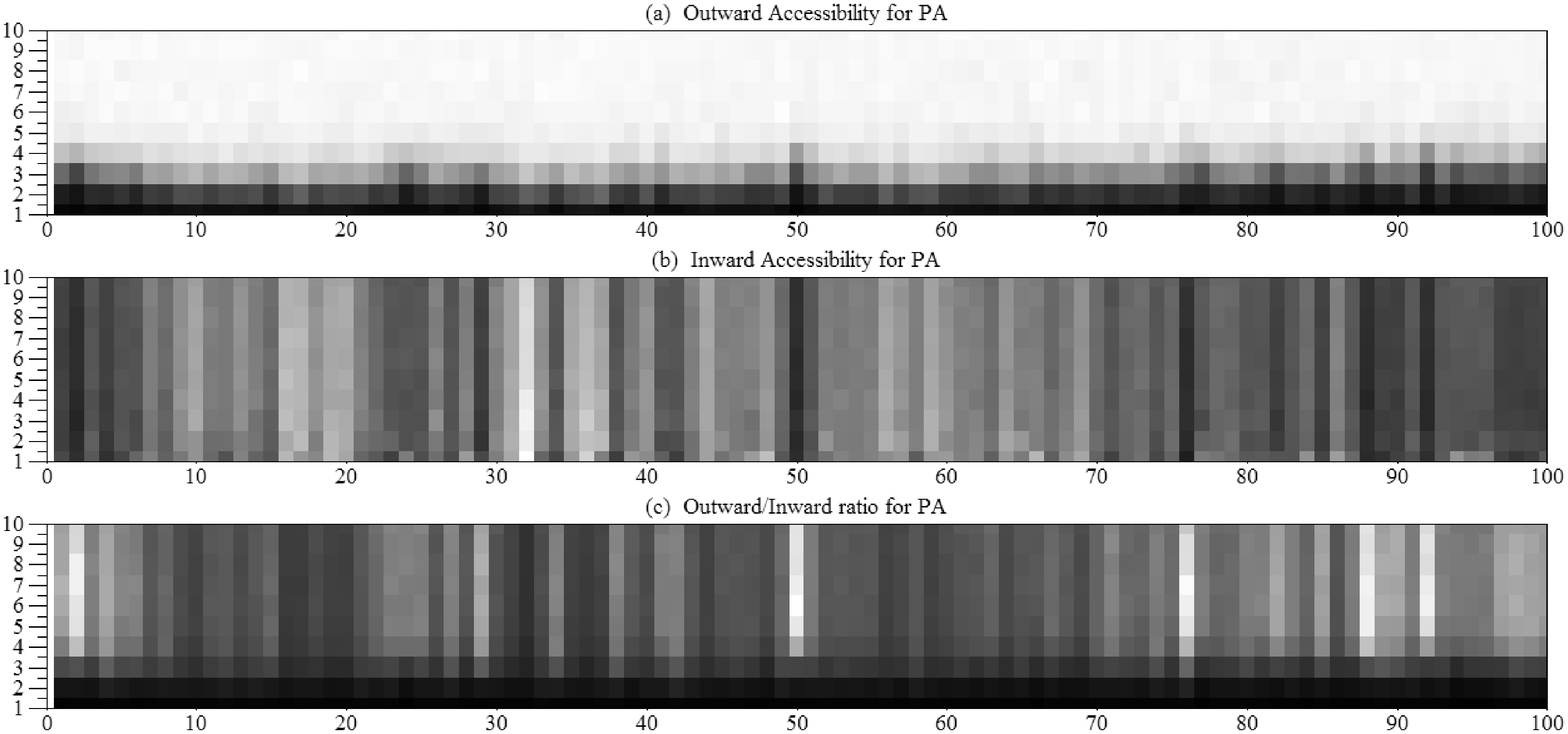}
  \caption{The outward accessibility (a), inward accessibility (b)
    and the outward/inward ratio (c) for the 10 initial steps 
    obtained for the considered PA network with $N=100$ nodes and
    $\left< k \right> = 6$.
    }~\label{fig:agram_PA}  \end{center}
\end{figure*}

Figures~\ref{fig:agram_ER} to ~\ref{fig:agram_PA} show the respective
measurements for the ER, BA, WS, GG, PN and PA networks. The $y-$axes
corresponds to the number of steps along the self-avoiding walks
(i.e. $h$).  The gray intensity in these images indicate higher values
of the measurements, normalized with respect to the maximum and
minimum values for the sake of better visualization.  A total of 2000
self-avoiding random walks were performed for each node by the
algorithm for numeric estimation of the transition probabilities
$P_h(i,j)$.

\begin{table*}
\centering
\begin{tabular}{|c||c|c|c|c|c|c|}  \hline  
   Network  &  \multicolumn{3}{|c|}{$OA$}   &  \multicolumn{3}{c|}{$IA$}  \\ \hline
& mean $\pm$ st. dev. & min & max & mean $\pm$ st. dev. & min  & max \\  \hline
     $ER$      &  $0.73 \pm 0.30$    &   0.001  &  0.73
               &  $0.010 \pm 0.003$  &   0.001  &  0.027       \\ \hline     
     $BA$      &  $0.64 \pm 0.26$    &   0.02   &  0.26
               &  $0.010 \pm 0.008$  &   0.001  &  0.096       \\ \hline     
     $WS$      &  $0.52 \pm 0.28$    &   0.03   &  0.94
               &  $0.010 \pm 0.001$  &   0.004  &  0.016       \\ \hline     
     $GG$      &  $0.28 \pm 0.17$    &   0.001  &  0.76
               &  $0.010 \pm 0.004$  &   0.002  &  0.035       \\ \hline     
     $PN$      &  $0.77 \pm 0.32$    &   0.50   &  0.98
               &  $0.010 \pm 0.001$  &   0.008  &  0.011       \\ \hline     
     $PA$      &  $0.71 \pm 0.31$    &   0.20   &  0.94
               &  $0.010 \pm 0.003$  &   0.003  &  0.023       \\ \hline     
\end{tabular}
\caption{The mean, standard deviations, minimum and maximum values 
         of the outward and inward
         accessibilities obtained for each of the networks, considering
         all the 10 steps.  Because of the normalization of the transition
         probabilities, the mean of $IA$ is always equal to $0.01$.
        }\label{tab:avst}
\end{table*}

The first important result is that the inward accessibility values
were verified to be much smaller than the respective outward
accessibilities, as can be verified from Table~\ref{tab:avst}.
Observe that the mean $IA$ values are necessarily $0.001 = 1/N$
because of the normalization of the probabilities.  Part of the
reasons for the asymmetry of accesses shown in this table is readily
clear from the example in Figure~\ref{fig:ex}(a): after two steps
along the walks, any of the surrounding nodes will have $OA_2 = 6/7$
and $IA_2 = 1/7$.  In this case, the marked asymmetry is caused by the
difference of degrees between the central and the surrounding nodes as
well as the fact that the surrounding nodes are attached to the
central node only.

\begin{figure}
  \vspace{0.3cm} \begin{center}
  \includegraphics[width=0.7\linewidth]{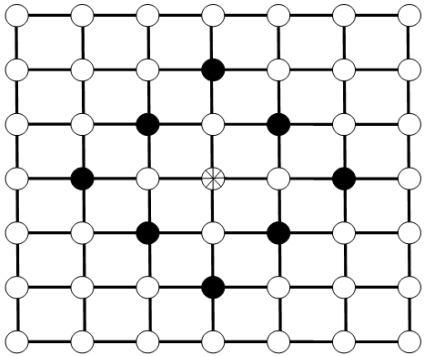} 
  \caption{A lattice network with $N=49$ nodes.
           The studied node (central) is marked with an asterisk,
           and the nodes which can be visited by the central node
           after two steps are shown in black.
    }~\label{fig:lattice}  \end{center}
\end{figure}

Because the regularity of networks is always a characteristic
affecting several of the topological properties, let us consider how
the accessibilities behave in a lattice network such as that shown in
Figure~\ref{fig:lattice}. Such a network is highly regular with
respect to the node degrees, except at its borders.  Let us study the
central node.  At step $h=1$, we have $OA_1 = 4/48$ and $IA_1 = 1/48$.
For $h=2$, we have $OA_2 = 8/48$ and $IA_2 = 1/48$. Observe that the
asymmetry tends to increase strongly with $h$.  Similar situations are
verified for the other nodes in the network. Note that in the case of
the network in Figure~\ref{fig:lattice}, the accessibilities to the
central node for $h=1$ and $h=2$ are not affected by border effects.

A little reasoning reveals that the marked asymmetry between the
outward and inward accessibilities in many networks is ultimately a
consequence of the degree regularity and high average degree, implying
that usually each node in a network tends to be connected to many
additional distinct nodes as $h$ increases.  Observe also a striking
difference between the situations defined by $P_1(central,surround)$
in the networks in Figure~\ref{fig:ex}(a) and~\ref{fig:lattice}: while
the surrounding nodes in the former case can only move to the central
node after one step (implying the central node to have maximum inward
accessibility), in the latter situation the nodes surrounding the
reference node can move to many other places (therefore decreasing
substantially the inward accessibility of the central node).  This
effect, which acts as a `diode' on the flow of accesses, tends to
increase substantially with $h$ and is the main cause of accessibility
asymmetry in typical complex networks.

\begin{figure*}
  \vspace{0.3cm} \begin{center}
  \includegraphics[width=0.45\linewidth]{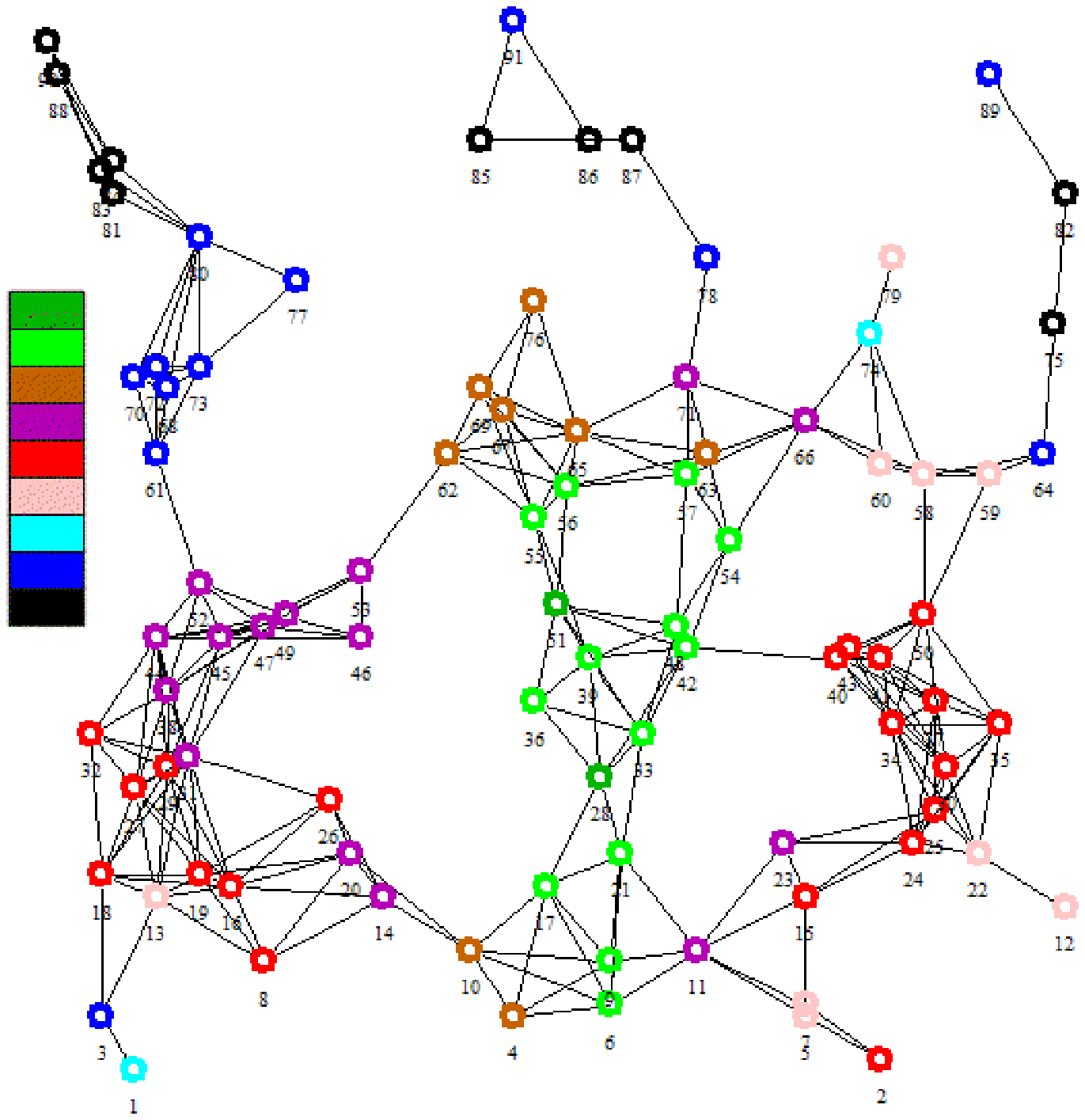} \hspace{0.1cm}
  \includegraphics[width=0.45\linewidth]{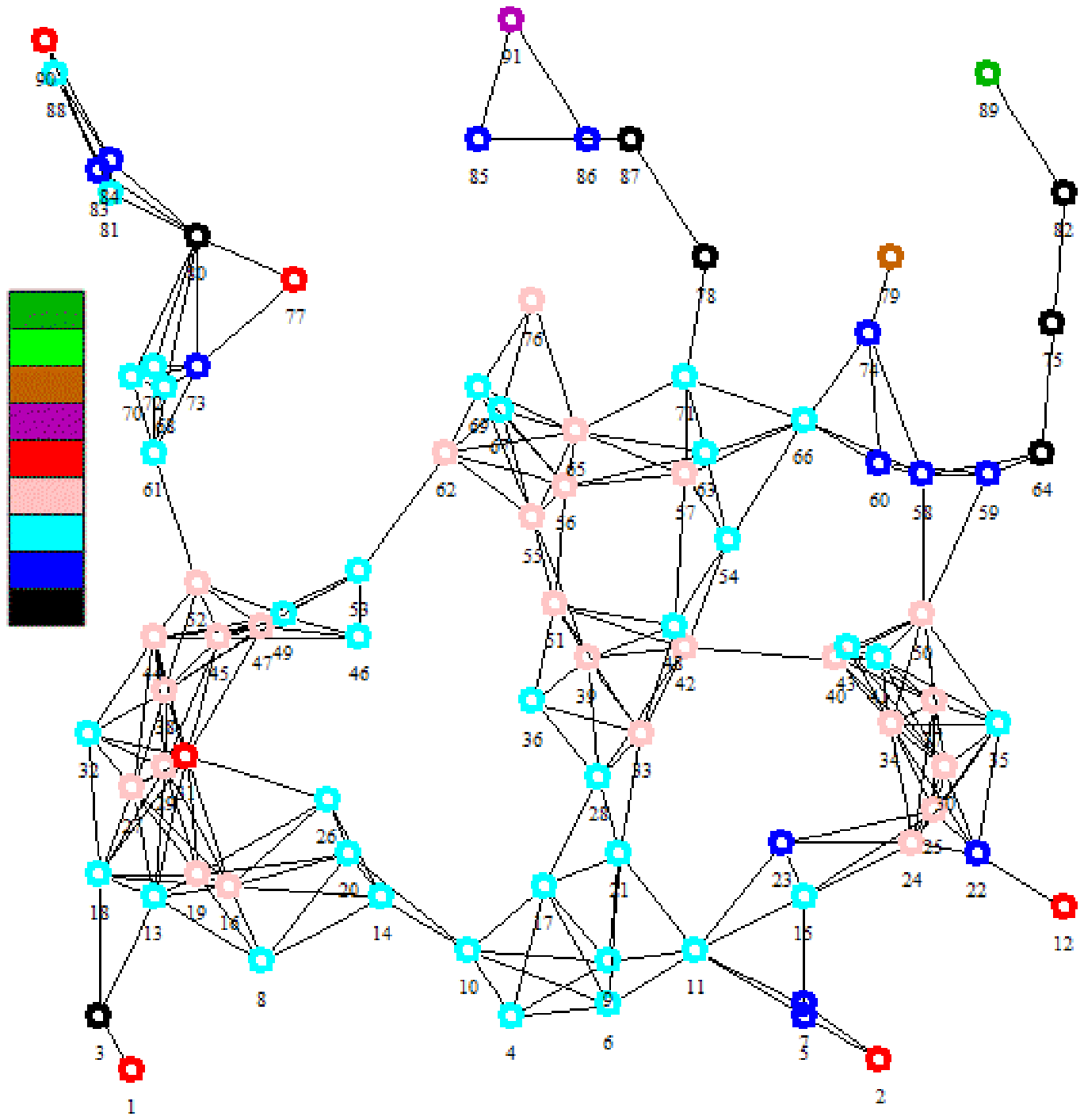} \\
  (a) \hspace{8cm} (b) \\
  \caption{The $OA$ (a) and $IA$ (b) accessibilities for the GG
    network considered in this article.  The color bar increases
    from bottom upwards.
    }~\label{fig:graphs}  \end{center}
\end{figure*}

Additional insights about the nature of asymmetries in specific cases
can be gathered by inspecting Figure~\ref{fig:graphs}, which show the
PCA outward (a) and inward (b) accessibilities for the considered GG
network.  These measurements were obtained by applying principal
component analysis (PCA)~\cite{Costa_book:2001, Costa_surv:2007}
considering all number of steps, so that the resulting principal
variables corresponded to the most relevant uncorrelated linear
combination of measurements.  First, observe that a group of nodes (in
green) at the center of the graph resulted with the largest $OA$
values (Figure~\ref{fig:graphs}(a)).  This is because such nodes have
more neighbors along the initial steps of the self-avoiding walks than
the nodes which resulted in smaller $OA$ values, which correspond to
the `border' of the graph.  Observe that the $OA$ values can therefore
be used to define the border of the network as corresponding to the
nodes with the smallest $OA$ values.  Now, let us analyse the $IA$
values shown in Figure~\ref{fig:graphs}(b).  An opposite tendency is
clearly observed in this case, with the `border' nodes yielding the
highest $IA$ values.  This is because such groups of nodes tend to act
as traps for the self-avoiding walks (recall that the moving agents
remain at the termination nodes).  Observe also that the nodes leading
to such traps (e.g. 3, 75 and 82) tend to have the smallest values of
$IA$, as they do not accumulate the agents.  The presence of traps,
i.e. nodes or group of nodes loosely connected with the rest of the
network, contributes greatly to decreasing the $IA$ values at the
other nodes.  In the specific case of the type of GG network
considered in this work, such border (in the topological sense) nodes
and groups of nodes tend to result at the spatial border of the graph.

A series of interesting results and insights can be inferred from
Figures~\ref{fig:agram_ER} to ~\ref{fig:agram_PA}, which shows the
evolution (bottom-up, along the vertical direction in the figures) of
the $OA$, $IA$ and outward/inward ratios along the 10 steps
($y-$axes) for each of the network nodes (along the $x-$axes).  We
discuss such results from the general to the specific perspective as
follows.  To start with, $OA$ increases with $h$ in all cases.  In
agreement with previous results considering the diversity entropy
signatures~\cite{Costa_diverse:2008}, the increase of $OA$ with $h$
tended to be more gradual for the WS and GG networks, possibly for the
same reasons identified in ~\cite{Costa_diverse:2008}, namely the fact
that pairs of nodes in these two types of networks tend to be less
adjacent regarding longer paths~\footnote{Here, the concept of
adjacency is used to reflect not only single-edge connectivity, but
also the paths of several lengths found between pairs of nodes. This
type of adjacency is henceforth called \emph{path-adjacency}.}.  $IA$
tended to remain nearly constant with $h$ for ER, WS, PN and PA.  This
is a surprising result, because it indicates that the inward
accessibility does not seem to depend on the length of the walks.  As
we will discuss later in this section, the $IA$ values are strongly
positively correlated with the individual node degree in most cases,
especially for $h \ge 3$.  In the case of GG (see
Figure~\ref{fig:agram_GG}(b)), the $IA$ values tended to increase,
decrease or to present a peak along the steps $h$.  Such a marked
difference between this model and the 5 other types of networks
suggests that the invariance of $IA$ with $h$ in the latter models
seems to be related to some intrinsic feature of the GG type of
network, possibly its strictly local node adjacency and lack of
bypasses shortening the distance between nodes, so that the $IA$
values in this case are more dependent on the local connectivity
structure of the network.

As shown in Figure~\ref{fig:agram_BA}, the $IA$ in the case of the BA
network was dominated by the hub corresponding to node 2, which has
degree 38.  Interestingly, the $IA$ values tended to oscillate with
$h$ for this node, decreasing for the largest values of $h$.  Another
interesting result regards the fact that $OA$ tended to get more
uniform among the nodes for the highest values of $h$, except for the
GG model.  This is a consequence of the fact that the $OA$ values tend
to increase steeply for most models (except WS and GG) towards
plateaux of similar heights for most nodes in the networks.
Therefore, after 2 or 3 time steps, most nodes in the ER, BA, WS, PN
and PA will tend to access a large portion of the network nodes.  The
long range connections and path-adjacencies in these networks seem to
play an important role in this effect.  In the case of the $GG$
structure, some nodes (e.g. nodes 4, 6, 8, 9, etc.) resulted with
markedly higher values of $OA$ along the highest values of $h$ -- see
Figure~\ref{fig:agram_GG}.  This implies that some nodes in the
network acquire an especially important role in accessing nodes as the
number of steps increase in this type of networks.

Unlike $OA$, the values of $IA$ tend to vary little with $h$ for each
node, except for the WS and PN cases, in the sense that substantial
similarity is observed among the $IA$ values of several nodes.  As a
matter of fact, $IA$ becomes intensely uniform for the largest values
of $h$ for the PN network (see Figure~\ref{fig:agram_PN}),
corroborating the enhanced uniformity of this type of network (see
also~\cite{Costa_comp:2007, Costa_longest:2007, Costa_diverse:2008}).
The few nodes in Figure~\ref{fig:agram_PN} presenting smaller values
of $IA$ have been verified to correspond to the initial and final
nodes of the self-avoiding walks which are used to construct that type
of network~\cite{Costa_comp:2007}.

Because the inward accessibilities are largely smaller than the
respective outward counterparts, the values of the ratios
outward/inward tended to mostly reflect the latter measurement, except
mainly for the PN case.  More specifically, the outward/inward ratios
in this network indicates that the extremity nodes used to build this
type of network tend to present a tendency to access more than being
accessed, a trend which increases with $h$.

Having investigated the overall properties of the inward and outward
accessibilities for examples of the 6 considered models, we now turn
our attention to the particularly relevant problem of trying to
predict the dynamical property of accessibility by considering
exclusively topological features of the networks.  Observe that this
issue is intrinsically related to the structure-dynamics paradigm in
complex networks research.  Though such an analysis can be performed
while considering several network measurements~\cite{Costa_surv:2007},
in the current article we focus our attention on correlations between
the individual node accessibility and the respective degree $k$.  A
similar investigation targeting the correlation between linear
diffusion (simulated by traditional random walks) and node degree has
been reported in~\cite{Costa_corrs:2007}.  That work showed that
though full positive correlation is verified between visits received
by nodes and the respective degree in the case of undirected,
connected networks, such a correlation is generally broken in the case
of directed networks.  A sufficient condition, namely that the in and
outdegree of all nodes be identical, was identified in that work.  The
consideration of non-linear dynamics implemented by self-avoiding
networks in the current article naturally breaks the full correlation
between node accessibility and degree.  Yet, a strong positive
correlation between the inward accessibility and node degree has been
verified for several of the 6 networks studied in the current work,
especially for $h$ larger than 3 or 4 steps.

\begin{figure*}
  \vspace{0.3cm} \begin{center}
  \includegraphics[width=0.95\linewidth]{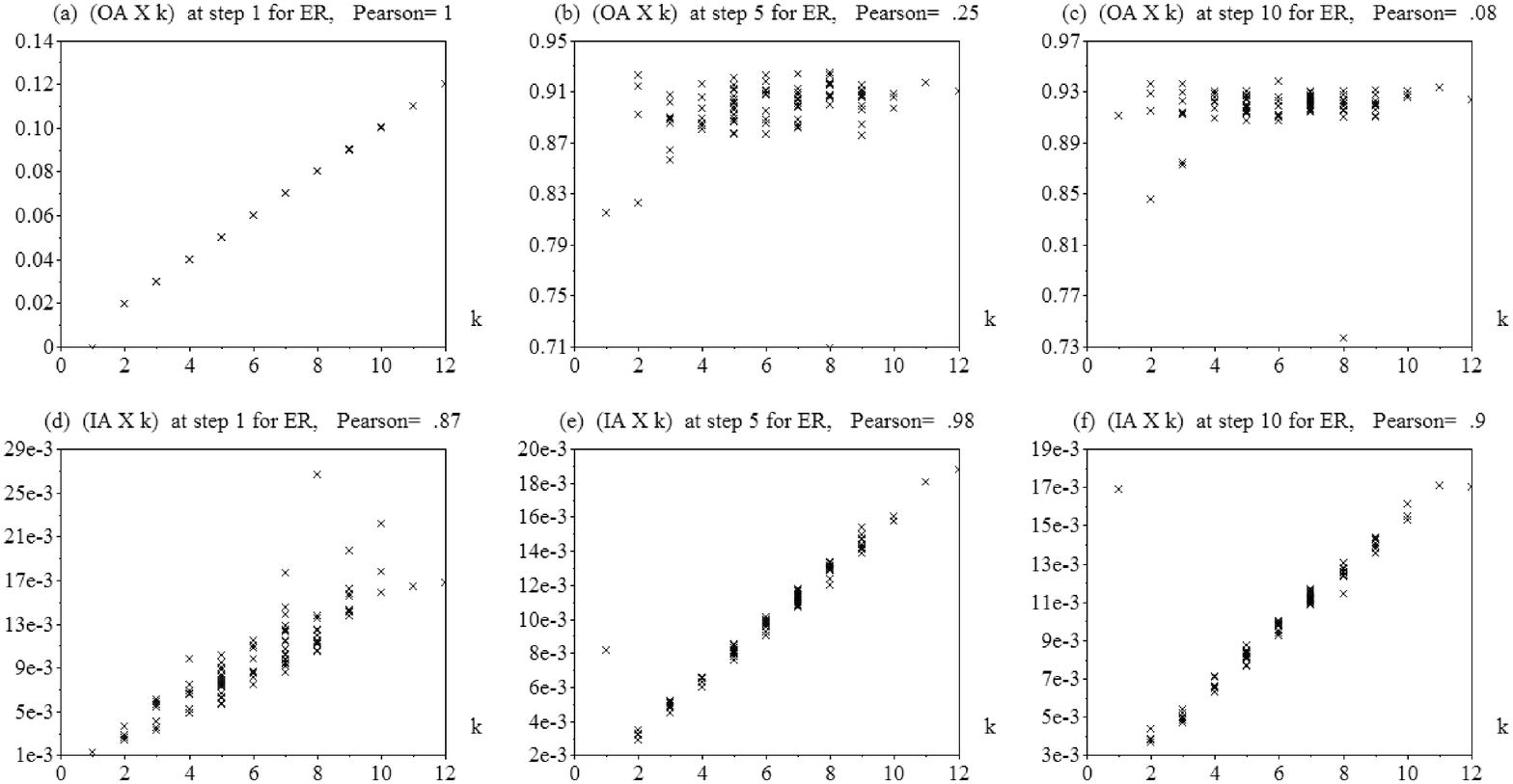} \caption{The
  scatterplots of the outward and inward accessibilities agains the
  node degree in the ER network, with respective Pearson correlation
  coefficients.}~\label{fig:corrs_ER} \end{center}
\end{figure*}

\begin{figure*}
  \vspace{0.3cm} \begin{center}
  \includegraphics[width=0.95\linewidth]{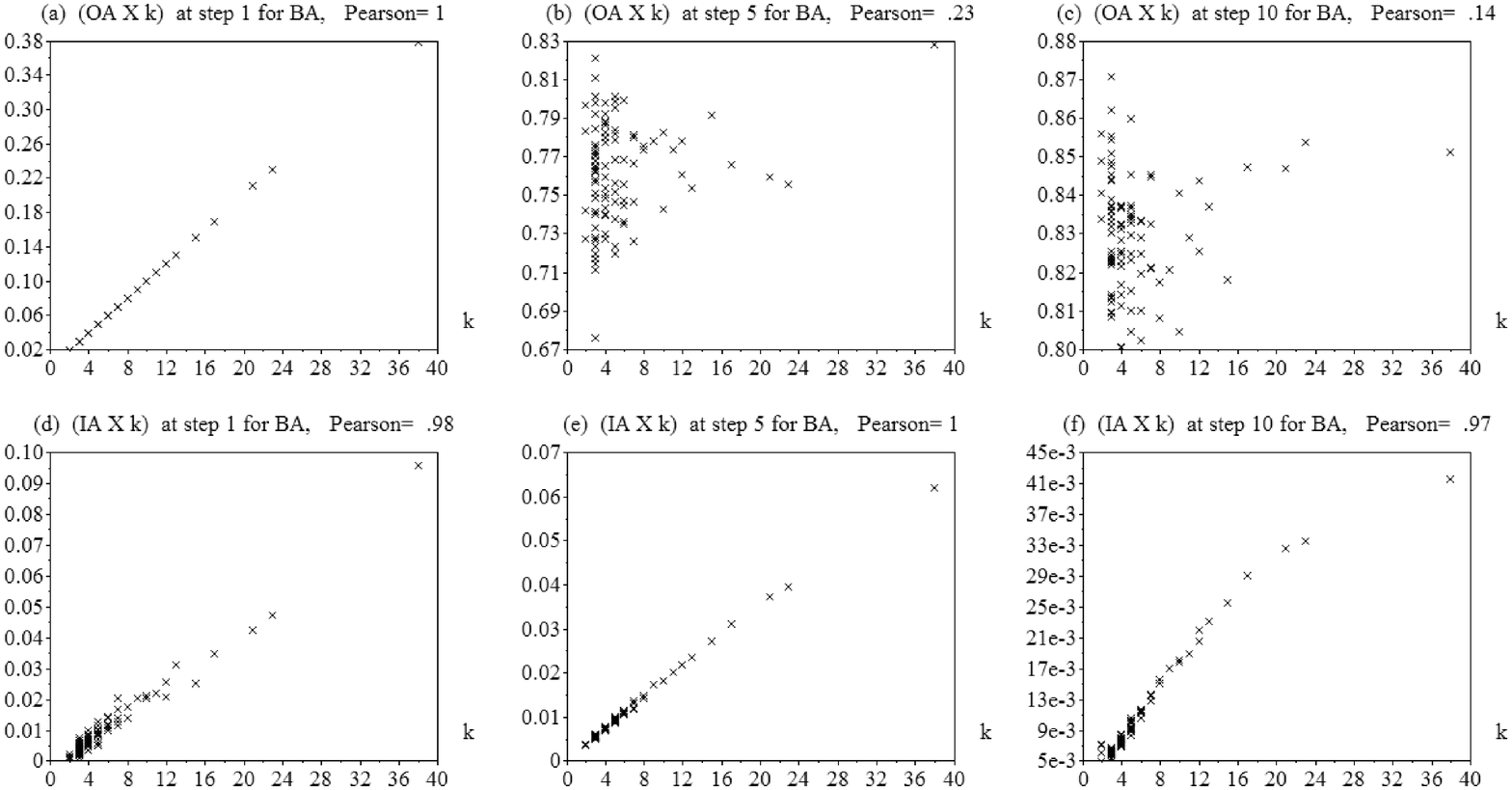}
  \caption{The scatterplots of the outward and inward accessibilities
    against the node degree in the BA network, with respective
    Pearson correlation coefficients.}~\label{fig:corrs_BA}  \end{center}
\end{figure*}

\begin{figure*}
  \vspace{0.3cm} \begin{center}
  \includegraphics[width=0.95\linewidth]{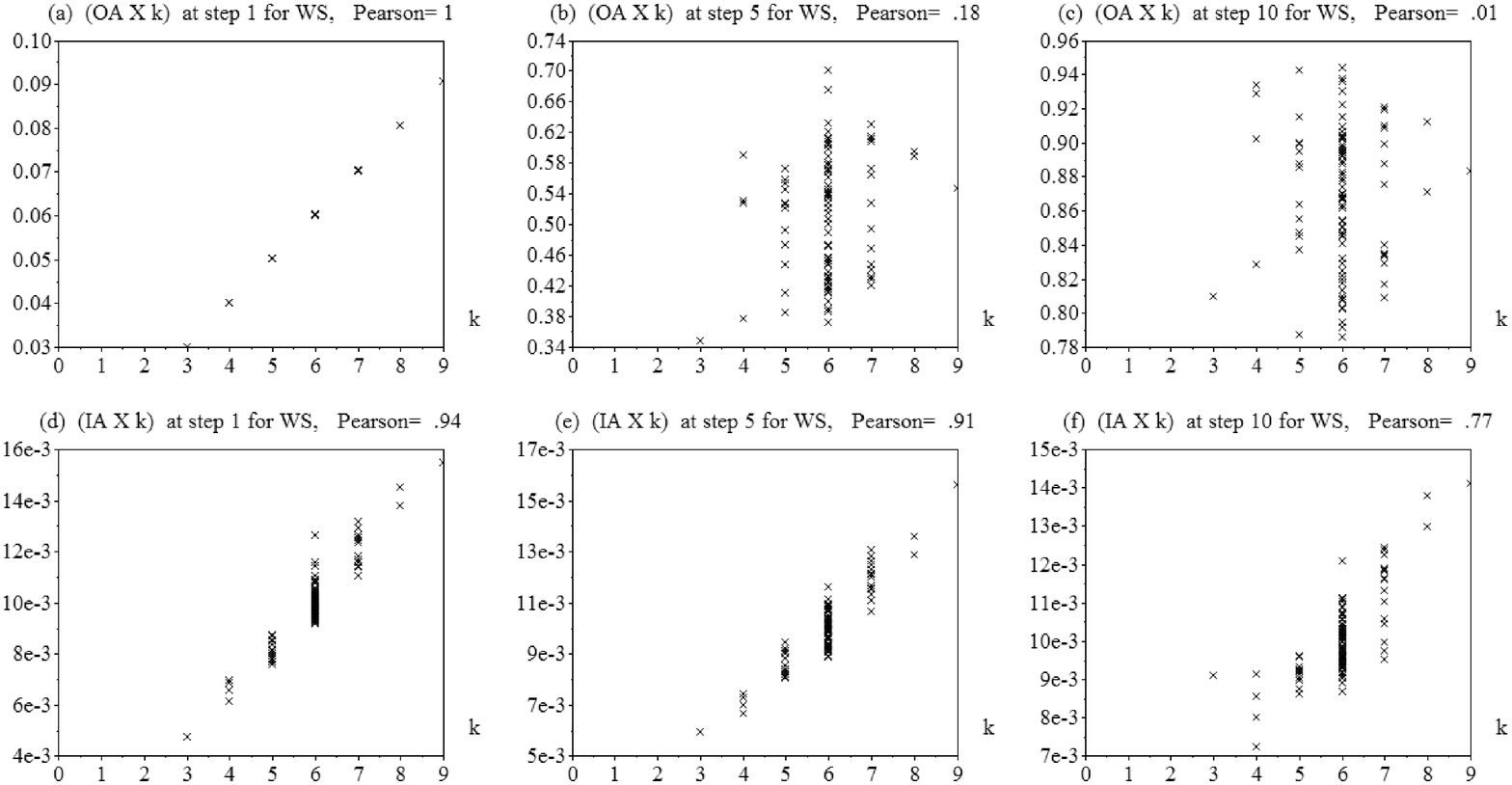}
  \caption{The scatterplots of the outward and inward accessibilities
    against the node degree in the WS network, with respective
    Pearson correlation coefficients.}~\label{fig:corrs_WS}  \end{center}
\end{figure*}

\begin{figure*}
  \vspace{0.3cm} \begin{center}
  \includegraphics[width=0.95\linewidth]{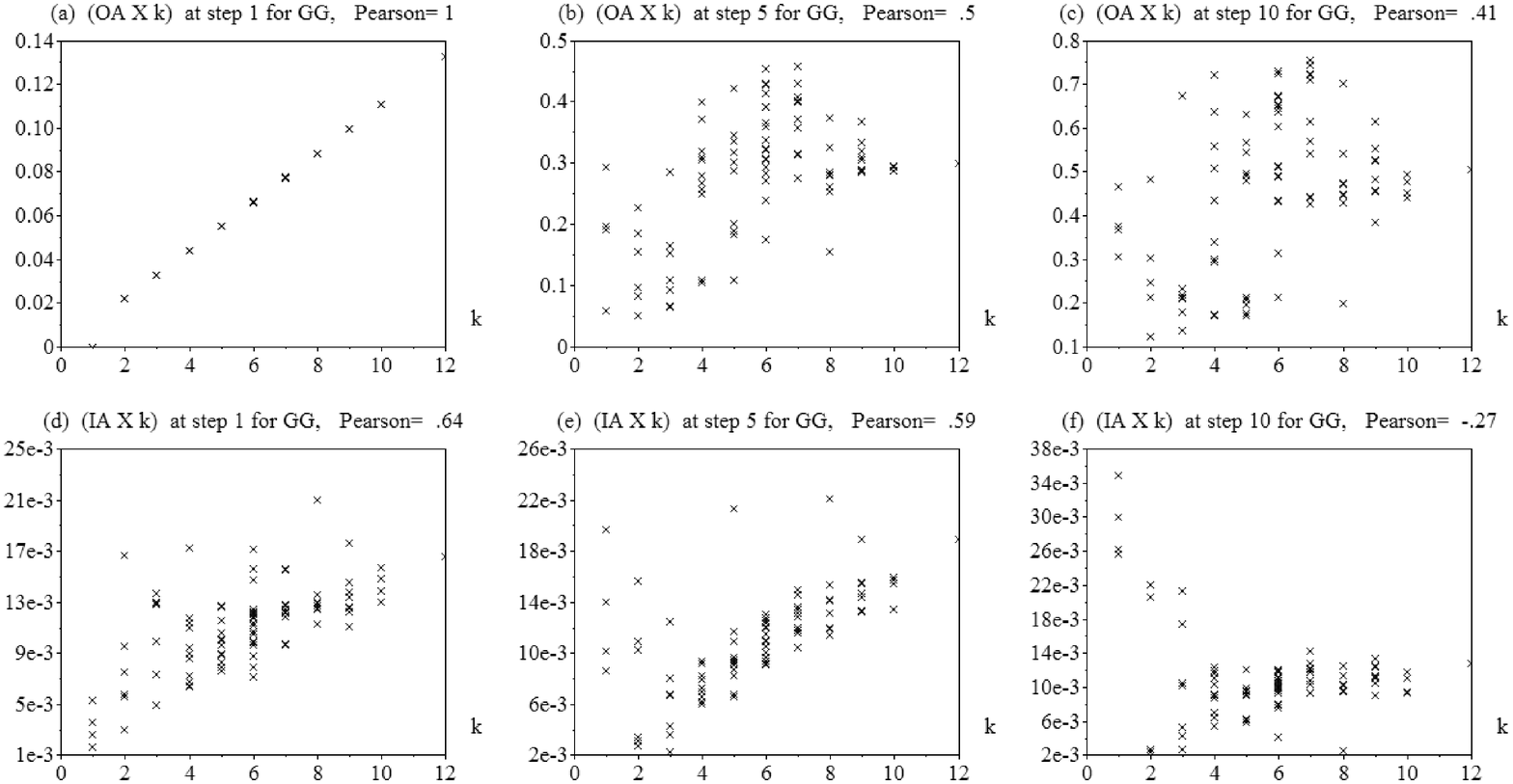}
  \caption{The scatterplots of the outward and inward accessibilities
    against the node degree in the GG network, with respective
    Pearson correlation coefficients.}~\label{fig:corrs_GG}  \end{center}
\end{figure*}

\begin{figure*}
  \vspace{0.3cm} \begin{center}
  \includegraphics[width=0.95\linewidth]{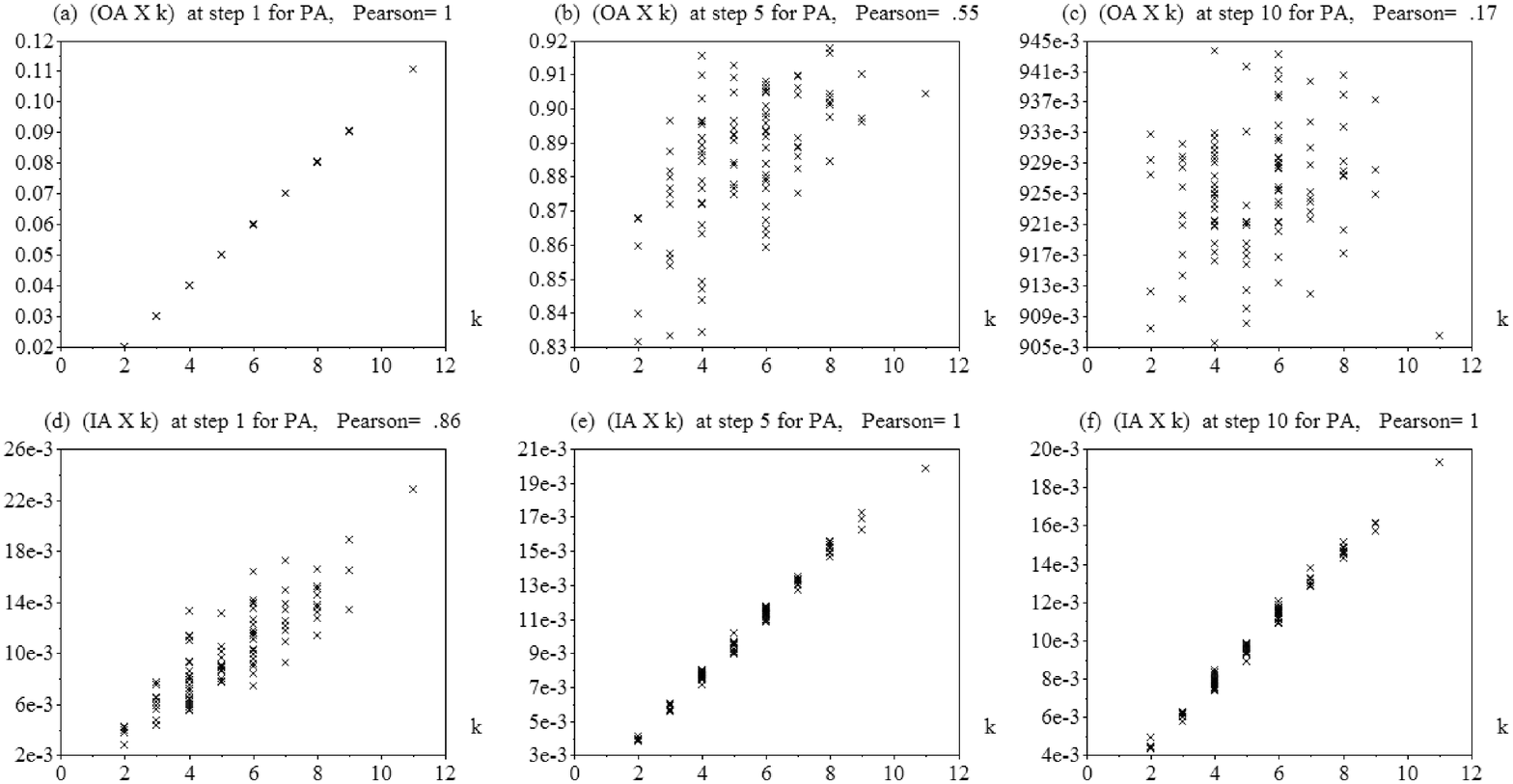}
  \caption{The scatterplots of the outward and inward accessibilities
    agains the node degree in the PA network, with respective
    Pearsotn correlation coefficients.}~\label{fig:corrs_PA}  \end{center}
\end{figure*}

\begin{figure*}
  \vspace{0.3cm} \begin{center}
  \includegraphics[width=0.95\linewidth]{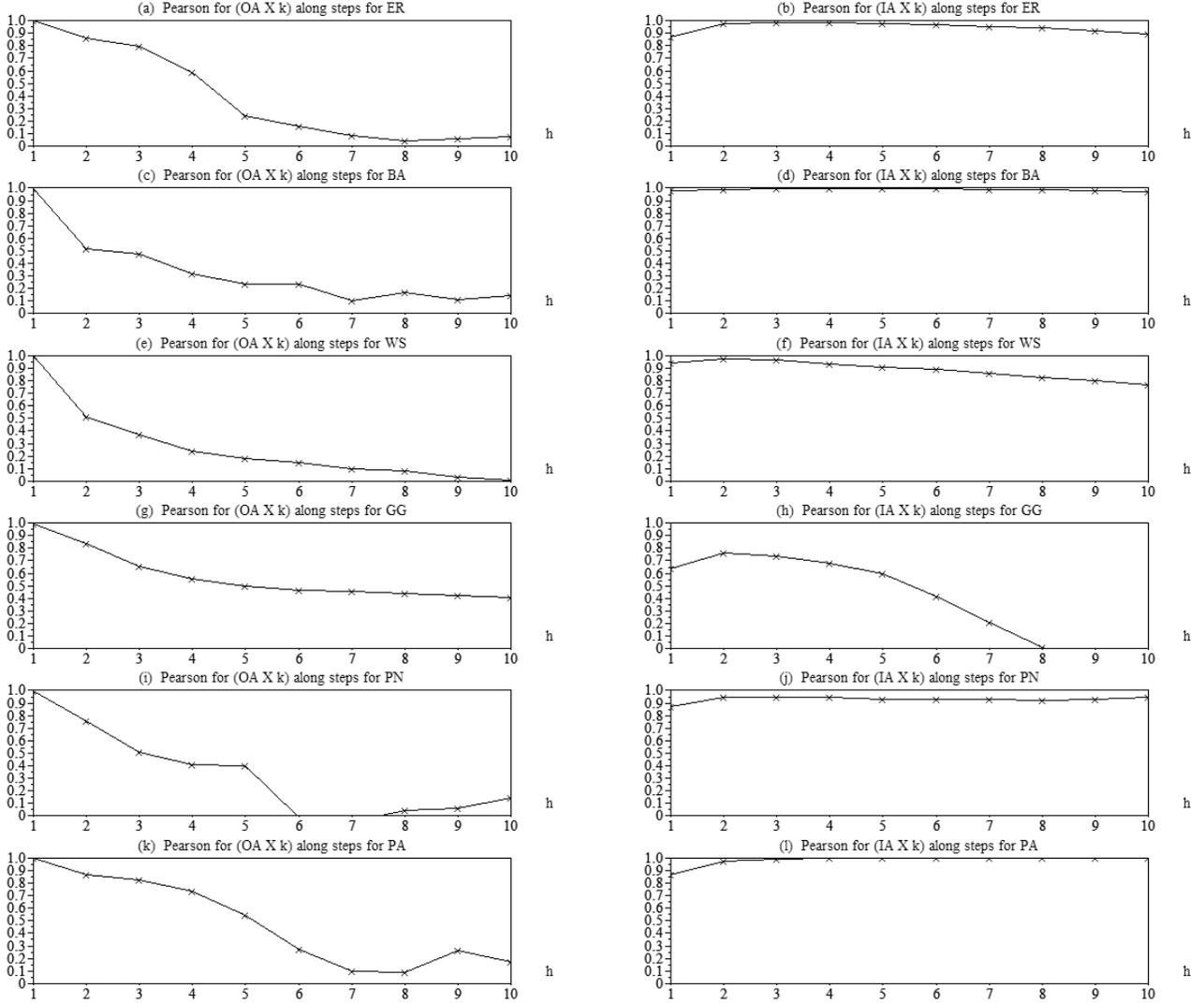} \caption{The
  Pearson correlation coefficients obtained between the outward and
  inward accessibilities and the node degrees for each of the
  considered networks.}~\label{fig:Pearsons} \end{center}
\end{figure*}

Figures~\ref{fig:corrs_ER} to~\ref{fig:corrs_PA} illustrate the scatterplots
of $OA \times k$ and $IA \times k$ respectively to the ER, BA, WS, GG,
PN and PA networks and $h=1$ (the initial step), $5$ (the middle step)
and 10 (the final steps in our self-avoiding walks).
Figure~\ref{fig:Pearsons} shows the evolution of the Pearson
correlation coefficients for the correlations $IA \times k$ and $OA
\times k$ in terms of $h$ for all networks.  Recall that the Pearson 
correlation coefficient varies from -1 (extreme negative correlation)
to 1 (extreme most positive correlation), with null value indicate
that the two measurement are uncorrelated.

\begin{figure*}
  \vspace{0.3cm} \begin{center}
  \includegraphics[width=0.95\linewidth]{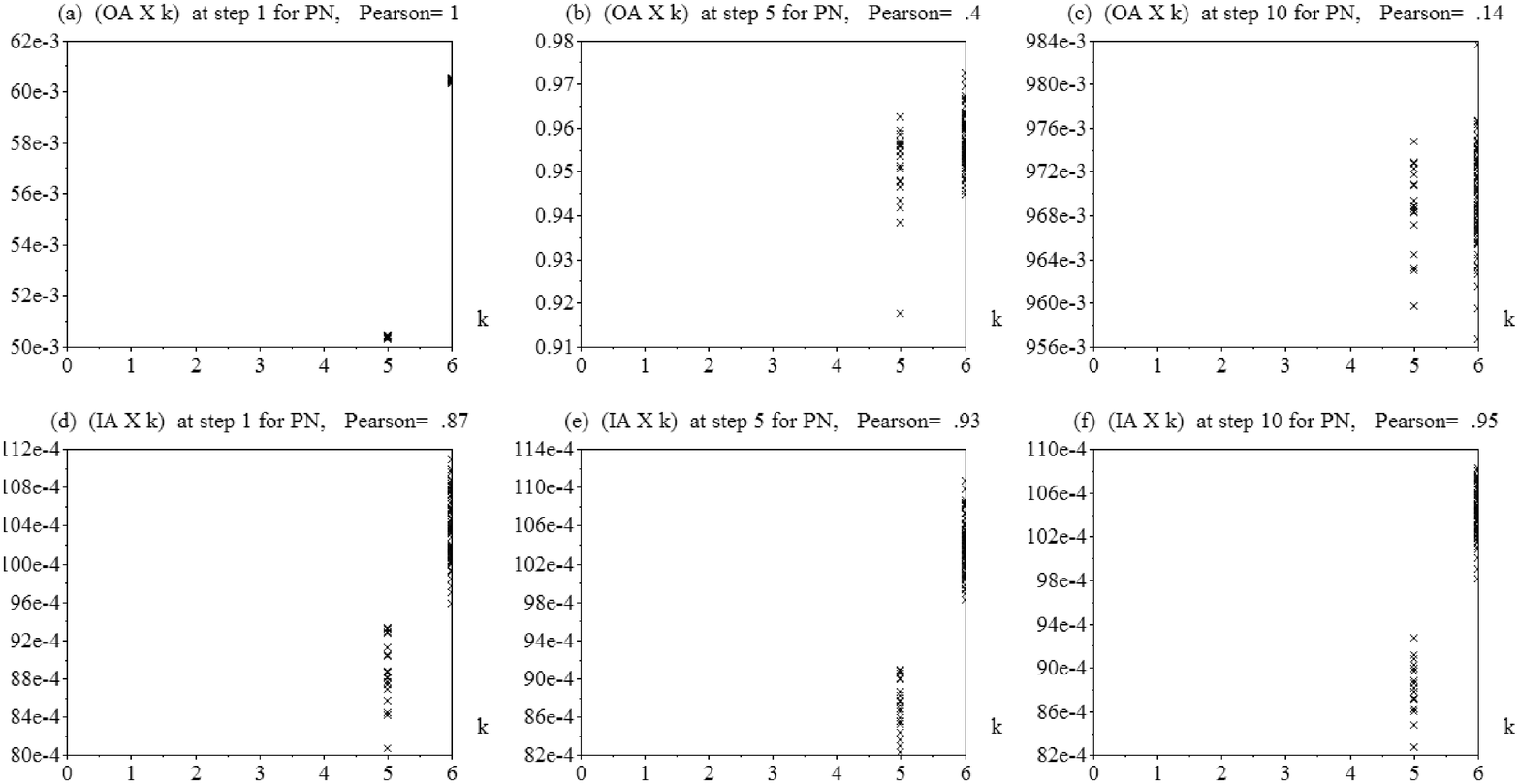}
  \caption{The scatterplots of the outward and inward accessibilities
    against the node degree in the PN network, with respective
    Pearson correlation coefficients.}~\label{fig:corrs_PN}  \end{center}
\end{figure*}

The first important conclusion regarding the relationship between the
outward accessibility and the individual node degree concerns the fact
that a full positive correlation is obtained between these two
features for $h=1$ (see Figures~\ref{fig:agram_ER}
to~\ref{fig:agram_PA} and Figure~\ref{fig:Pearsons}).  However, such a
correlation decreases steeply with $h$, and becomes almost irrelevant
after 4 or 5 steps (see Figure~\ref{fig:Pearsons}).  These results
imply that the outward accessibility of each node can only be used to
predict the respective outward accessibility for the 3 or 4 initial
steps.

Interestingly, an opposite trend has been observed for the correlation
between the inward accessibility and the degree, which is generally
high at $h=1$ and tends to increase further with $h$ for most cases,
except WS and GG (see Figure~\ref{fig:Pearsons}).  This result implies
that the inward accessibility of a node can be predicted with
remarkable accuracy from the respective node degree, especially for
relative high values of $h$.  This is surprising because the higher
the value of $h$, the more global the dynamics is unfolded along the
topology of the network.  The surprise therefore resides in the fact
that the node degree, an intrinsically local property of the network
connectivity, can be used to predict accurately (except for WS and GG)
the global transient dynamics of the self-avoiding walks for most
values of $h$.  Such a result has particularly important practical
implications.  For instance, in a system properly modeled by the
self-avoiding walks unfolding on complex networks similar to ER, BA,
PN and PA, it is possible to predict the number of visits to each node
in terms of its degree.

The above result has major implications for several real-world
problems.  For instance, in the case of disease spreading or node
attack, the more susceptible nodes can be easily identified and
protected.  If the dynamics is used to model neuronal or dissemination
of cortical activation, the more active nodes can be immediately
identified from their degrees.  The strong correlation between the
inward accessibility and the individual node degree also paves the way
for interfering in the local network topology (e.g. changing the
degree of specifically critical nodes) in order to address specific
needs.  For instance, in the case of WWW surfing, a site (i.e. a node)
can get more accessed provided it can motivate additional links
(however, note that this type of network would typically be directed,
so that such an application would require additional studies on
self-avoiding walks on that type of asymmetric connectivity).  The
correlation between inward accessibility and degree verified for most
of the networks is strongly related to the fact that, once the walks
have become spread along the network as $h$ increases (the ER, BA, PN
and PA networks tend to have strong and regular path-adjacency), most
edges tend to exhibit similar rates of accesses, so that the final
inward accessibility is ultimately defined by the individual node
degree, which taps visits with an intensity which is almost linear
with the number of incoming edges.  The strategy of enhancing the $IA$
correlation by adding new connections with other nodes will not
generally work in the GG case because in that case a node typically
receives connections only from nearby nodes (regarding spatial
distance), so that it will tap accesses from nodes with similar ---
and possibly low --- accessibility.  Ultimately, it is such a
possibility of segregation of levels of accessibility which
contributes to the break of the $IA \times k$ correlation in the case
of the GG (and also to a certain extent in the WS) model.

Interestingly, the importance of uniformity of edge activation for the
inward correlation also explains why it is so difficult to predict the
outward accessibility from the degree larger values of $h$: the
problem is that the dissemination of the accesses tends to become
uniform as the walks unfold along longer lengths irrespectively of the
local connectivity along the initial steps of the walks.  These
results and interpretations suggest investigations of the
accessibility to \emph{edges}, in addition to the studies targeting
nodes reported in the current work.

Still considering the positive correlation between the inward
accessibility and degree, It can be verified from
Figure~\ref{fig:agram_BA} that the correlation in the scatterplot for
$h=10$ suggests a saturation of that relationship, manifested by the
sigmoid bending at the right-hand side of the curve.  Another
interesting issue regards the reasons for the lack of correlation
between the inward accessibility and the degree observed for the GG
model.  Strictly speaking, some significant correlation is observed
for $h=1$ to 3 or 4 (see Figure~\ref{fig:Pearsons}), but it tends to
fade away for larger values of $h$.  One possible explanation is that
the more local adjacency between nodes in this type of network (a
consequence of spatial adjacency, which tends to present communities
for the considered size --- see Figure~\ref{fig:graphs}), implies that
the flow of accesses along the edges of the network for large values
of $h$ will not be so uniform, making it impossible to predict the
inward accessibility in terms of the individual node degrees.

If the GG network implied the worst predictability, the PA allowed the
opposite feature, exhibiting Pearson correllation coefficient for $IA
\times k$ very close to 1 except for $h=1$.  This result seems to be 
related to the fact that paths of several lengths are guaranteed to
exist among several of the nodes in this type of networks as a
consequence of the way in which they are generated, i.e. by
transforming the star connectivity into path connectivity.  

\section{Concluding Remarks}

One particularly interesting means to investigate the relationship
between the dynamics and structure in complex networks involves the
study of non-linear random walks such as the self-avoiding walks.
Unlike traditional walks, that type of walk is more purposive and
effective in the sense of economically visiting nodes, i.e. with the
minimal possible number of edges.  Because such a kind of walk cannot
repeat nodes or edges, it implies in a more effective diffusion of
visits away from the original node, without the returns implied by the
traditional random walks.  In addition, unlike their traditional
counterparts, self-avoiding walks are necessarily finite.  Despite
such important and useful features, self-avoiding walks have received
relatives little attention from the complex networks literature.  Even
less frequently studied is the transient dynamics in such systems.

Another important aspect regarding the study of dynamics in complex
networks regards the accessibility of a node.  This property is
critical because it underlies a series of important theoretical and
real-world problems, including disease spreading, WWW navigation,
protein interaction, transportation, urban planning, communications,
distributed computing, cortical network activity, neuronal networks
(where the refractory period of neuronal firing can be modeled by the
self-avoiding criterion), amongst many other cases.

The current article has brought together these two relevant issues,
namely self-avoiding walks and node in ward and outward accessibility.
More specifically, we build up from previous
results~\cite{Costa_diverse:2008} in order to propose two sound and
meaningful definitions of the accessibilities.  Therefore, by
\emph{outward accessibility} we quantify the potential of each node in
a given network to be accessed by self-avoiding walks originating at
all other network nodes.  The
\emph{inward accessibility} has an analogue meaning regarding the accesses 
received by each specific node.  Both such measurements can be
calculated from the transition probabilities between any pair of
distinct nodes for each considered number of steps.  In this work,
these probabilities have been estimated by using a simple algorithm
which consists exactly in implementing the considered dynamics,
i.e. performing a large number of self-avoiding walks from each node.
The particularly relevant issue of trying to predict the dynamical
property of accessibility by considering exclusively topological
features of the networks (in the case of the present work the
individual node degree) has also been addressed. This issue is at the
very center of the structure/dynamics relationship paradigm.

Because of the quantity of results, the reader is kindly requested to
refer to Section~\ref{sec:results} for a complete listing and
discussion of the results.  Only the most relevant results and
interpretations are listed in the following.  The first result which
stands out regards the fact that the outward accessibility is
invariably much larger than the inward accessibility for most nodes.
This has been verified to be a consequence of the relatively large
degrees and regularity observed for most of the considered networks,
which enhances the possibilities that the outgoing walks diverge among
the network.  As a matter of fact, observe that the outward
accessibility presents an analogy with the Lyapunov coefficient in
dynamics systems, in the sense that it quantifies the divergence of
the dynamics along the consecutive steps of the walks.  Another
particularly relevant result obtained in this work concerns the fact
that most accessibility values for each individual node tend to
increase with $h$.  A series of results and insights were additionally
obtained regarding the application of the measurements to 6 distinct
types of complex networks, namely Erd\H{o}s-R\'enyi,
Barab\'asi-Albert, Watts-Strogatz, a simple geographical type of
network, as well as two knitted networks (PN and PA).  Out of these
models, the GG case led to the most different dynamics, as a
consequence of its intense local adjacency.  The accessibilities were
found to increase more abruptly along $h$ for the ER, BA, PN and PA
structures.  Except for the GG model, the outward accessibility was
moderately correlated with the node degree along the initial steps for
all models, fading away after 3 or 4 steps. Surprisingly, the inward
accessibility tended to be high and to increase further with the number
of steps (except of the WS and GG case), implying that such a global
dynamical property can be accurately predicted from the degree of
individual nodes.  This important result is related to the
uniformization of the flow accesses among the edges typically achieved
for larger values of $h$.

Several are the possibilities for future investigations motivated by
the currently reported concepts, methods and results.  To begin with,
it would be interesting to consider other values of average degree and
network sizes, though this will require considerable computational
resources.  Because of the intrinsic importance of accessibility, it
would be particularly interesting to characterize the inward and
outward accessibility of real-world networks, such as those related to
disease spreading, transportation, biological interactions, amongst
many others.  Also, as the definition of accessibilities proposed in
this article are independent of the type of network and random walk,
it would be easy and immediate to extend such analyses to additional
theoretical network models and less conventional random walks such as
those described in~\cite{Costa_know:2006}, including direct and
weighted structures.  Particularly promising is the application of
random walks formulation to model and study synchronization in linear
and non-linear systems, e.g. by considering nodes with similar
effective period (estimated as the inverse of the respective
transition probabilities, which play a role of mean frequency of
accesses) of received visits.  Because of the correlations between the
accessibilities and the individual node degrees are to a large extent
related to the uniformity of access rates (or flows) along the edges,
it would be particularly interesting to investigate in a systematic
way the dynamics of the accesses along edges, extending the concept of
accessibility to edges in addition to nodes.  Another possibility
worth being pursued is to allow the walks to start only at specific
subsets of nodes, so that more specific aspects of the dynamics and
its relationship with specific connectivity patterns of the network
can be dissected.  Of particular interest among all the possibilities
defined by the current work is possibly to use the strong correlations
between the inward accessibility and node degree in order to make
predictions about the former in terms of the latter, allowing a series
of important applications to problems involving real-world networks.
Actually, the identification of such a close relationship between a
global dynamical feature and a inherently local topological property
provides a striking example of the importance that structure-dynamics
studies can have for complex networks research.

\begin{acknowledgments}
Luciano da F. Costa thanks CNPq (308231/03-1) and FAPESP (05/00587-5)
for sponsorship.
\end{acknowledgments}

\bibliography{nlin_diff}

\begin{thebibliography}{31}
\expandafter\ifx\csname natexlab\endcsname\relax\def\natexlab#1{#1}\fi
\expandafter\ifx\csname bibnamefont\endcsname\relax
  \def\bibnamefont#1{#1}\fi
\expandafter\ifx\csname bibfnamefont\endcsname\relax
  \def\bibfnamefont#1{#1}\fi
\expandafter\ifx\csname citenamefont\endcsname\relax
  \def\citenamefont#1{#1}\fi
\expandafter\ifx\csname url\endcsname\relax
  \def\url#1{\texttt{#1}}\fi
\expandafter\ifx\csname urlprefix\endcsname\relax\def\urlprefix{URL }\fi
\providecommand{\bibinfo}[2]{#2}
\providecommand{\eprint}[2][]{\url{#2}}

\bibitem[{\citenamefont{Newman}(2003)}]{Newman:2003}
\bibinfo{author}{\bibfnamefont{M.~E.~J.} \bibnamefont{Newman}},
  \bibinfo{journal}{SIAM Rev.} \textbf{\bibinfo{volume}{45}},
  \bibinfo{pages}{167} (\bibinfo{year}{2003}).

\bibitem[{\citenamefont{Boccaletti et~al.}(2006)\citenamefont{Boccaletti,
  Latora, Moreno, Chavez, and Hwang}}]{Boccaletti:2006}
\bibinfo{author}{\bibfnamefont{S.}~\bibnamefont{Boccaletti}},
  \bibinfo{author}{\bibfnamefont{V.}~\bibnamefont{Latora}},
  \bibinfo{author}{\bibfnamefont{Y.}~\bibnamefont{Moreno}},
  \bibinfo{author}{\bibfnamefont{M.}~\bibnamefont{Chavez}}, \bibnamefont{and}
  \bibinfo{author}{\bibfnamefont{D.}~\bibnamefont{Hwang}},
  \bibinfo{journal}{Phys. Rep.} \textbf{\bibinfo{volume}{424}},
  \bibinfo{pages}{175} (\bibinfo{year}{2006}).

\bibitem[{\citenamefont{Stauffer et~al.}(2003)\citenamefont{Stauffer, Aharony,
  da~F.~Costa, and Adler}}]{Stauffer_Hopfield}
\bibinfo{author}{\bibfnamefont{D.}~\bibnamefont{Stauffer}},
  \bibinfo{author}{\bibfnamefont{L.}~\bibnamefont{Aharony}},
  \bibinfo{author}{\bibfnamefont{L.}~\bibnamefont{da~F.~Costa}},
  \bibnamefont{and} \bibinfo{author}{\bibfnamefont{J.}~\bibnamefont{Adler}},
  \bibinfo{journal}{Eur. Phys. J. B} \textbf{\bibinfo{volume}{32}},
  \bibinfo{pages}{395} (\bibinfo{year}{2003}).

\bibitem[{\citenamefont{da~F.~Costa and Stauffer}(2003)}]{Stauffer_Costa}
\bibinfo{author}{\bibfnamefont{L.}~\bibnamefont{da~F.~Costa}} \bibnamefont{and}
  \bibinfo{author}{\bibfnamefont{D.}~\bibnamefont{Stauffer}},
  \bibinfo{journal}{Physica A} \textbf{\bibinfo{volume}{330}},
  \bibinfo{pages}{37} (\bibinfo{year}{2003}).

\bibitem[{\citenamefont{Aleksiejuk et~al.}(2002)\citenamefont{Aleksiejuk,
  Holyst, and Stauffer}}]{Ising_BA}
\bibinfo{author}{\bibfnamefont{A.}~\bibnamefont{Aleksiejuk}},
  \bibinfo{author}{\bibfnamefont{J.~A.} \bibnamefont{Holyst}},
  \bibnamefont{and} \bibinfo{author}{\bibfnamefont{D.}~\bibnamefont{Stauffer}},
  \bibinfo{journal}{Physica A} \textbf{\bibinfo{volume}{310}},
  \bibinfo{pages}{260} (\bibinfo{year}{2002}).

\bibitem[{\citenamefont{Stauffer and Kulakowski}(2002)}]{Ising_BA2}
\bibinfo{author}{\bibfnamefont{D.}~\bibnamefont{Stauffer}} \bibnamefont{and}
  \bibinfo{author}{\bibfnamefont{K.}~\bibnamefont{Kulakowski}}
  (\bibinfo{year}{2002}), \bibinfo{note}{arXiv:cond-mat/0210225}.

\bibitem[{\citenamefont{da~F.~Costa and Sporns}(2007)}]{Costa_Ising}
\bibinfo{author}{\bibfnamefont{L.}~\bibnamefont{da~F.~Costa}} \bibnamefont{and}
  \bibinfo{author}{\bibfnamefont{O.}~\bibnamefont{Sporns}},
  \bibinfo{journal}{Intl. J. Bif. Chaos} \textbf{\bibinfo{volume}{17}},
  \bibinfo{pages}{2387} (\bibinfo{year}{2007}).

\bibitem[{\citenamefont{Bernardes et~al.}(2002)\citenamefont{Bernardes,
  Stauffer, and Kertesz}}]{Sznajd_BA}
\bibinfo{author}{\bibfnamefont{A.~T.} \bibnamefont{Bernardes}},
  \bibinfo{author}{\bibfnamefont{D.}~\bibnamefont{Stauffer}}, \bibnamefont{and}
  \bibinfo{author}{\bibfnamefont{J.}~\bibnamefont{Kertesz}},
  \bibinfo{journal}{Eur. Phys. J. B} \textbf{\bibinfo{volume}{25}},
  \bibinfo{pages}{123} (\bibinfo{year}{2002}).

\bibitem[{\citenamefont{Fortunato}(2004)}]{Fortunato_opinion}
\bibinfo{author}{\bibfnamefont{S.}~\bibnamefont{Fortunato}}
  (\bibinfo{year}{2004}), \bibinfo{note}{arXiv:cond-mat/0405083}.

\bibitem[{\citenamefont{Zhou}(2003)}]{Zhou:2003}
\bibinfo{author}{\bibfnamefont{H.}~\bibnamefont{Zhou}}, \bibinfo{journal}{Phys.
  Rev. E} \textbf{\bibinfo{volume}{67}}, \bibinfo{pages}{061901}
  (\bibinfo{year}{2003}), \bibinfo{note}{arXiv:physics/0302032}.

\bibitem[{\citenamefont{Noh and Rieger}(2004)}]{Rieger:2004}
\bibinfo{author}{\bibfnamefont{J.~D.} \bibnamefont{Noh}} \bibnamefont{and}
  \bibinfo{author}{\bibfnamefont{H.}~\bibnamefont{Rieger}},
  \bibinfo{journal}{Phys. Rev. Letts.} \textbf{\bibinfo{volume}{92}},
  \bibinfo{pages}{118701} (\bibinfo{year}{2004}),
  \bibinfo{note}{arXiv:cond-mat/0307719}.

\bibitem[{\citenamefont{Masuda and Konno}(2004)}]{Masuda:2004}
\bibinfo{author}{\bibfnamefont{N.}~\bibnamefont{Masuda}} \bibnamefont{and}
  \bibinfo{author}{\bibfnamefont{N.}~\bibnamefont{Konno}},
  \bibinfo{journal}{Phys. Rev. E} \textbf{\bibinfo{volume}{69}},
  \bibinfo{pages}{066113} (\bibinfo{year}{2004}),
  \bibinfo{note}{arXiv:cond-mat/0401255}.

\bibitem[{\citenamefont{Pons and Latapy}(2005)}]{Pons_comm:2005}
\bibinfo{author}{\bibfnamefont{P.}~\bibnamefont{Pons}} \bibnamefont{and}
  \bibinfo{author}{\bibfnamefont{M.}~\bibnamefont{Latapy}}
  (\bibinfo{year}{2005}), \bibinfo{note}{arXiv:physics/0512106}.

\bibitem[{\citenamefont{Eisler and Kertesz}(2005)}]{Eisler:2005}
\bibinfo{author}{\bibfnamefont{Z.}~\bibnamefont{Eisler}} \bibnamefont{and}
  \bibinfo{author}{\bibfnamefont{J.}~\bibnamefont{Kertesz}},
  \bibinfo{journal}{Phys. Rev. E} \textbf{\bibinfo{volume}{71}},
  \bibinfo{pages}{057104} (\bibinfo{year}{2005}),
  \bibinfo{note}{arXiv:physics/0512106}.

\bibitem[{\citenamefont{da~F.~Costa
  et~al.}(2007{\natexlab{a}})\citenamefont{da~F.~Costa, Sporns, Antiqueira,
  Nunes, and Oliveira}}]{Costa_corrs:2007}
\bibinfo{author}{\bibfnamefont{L.}~\bibnamefont{da~F.~Costa}},
  \bibinfo{author}{\bibfnamefont{O.}~\bibnamefont{Sporns}},
  \bibinfo{author}{\bibfnamefont{L.}~\bibnamefont{Antiqueira}},
  \bibinfo{author}{\bibfnamefont{M.~G.~V.} \bibnamefont{Nunes}},
  \bibnamefont{and} \bibinfo{author}{\bibfnamefont{O.~N.}
  \bibnamefont{Oliveira}}, \bibinfo{journal}{Appl. Phys. Letts.}
  \textbf{\bibinfo{volume}{91}}, \bibinfo{pages}{054107}
  (\bibinfo{year}{2007}{\natexlab{a}}).

\bibitem[{\citenamefont{Kinouchi et~al.}(2002)\citenamefont{Kinouchi, Martinez,
  Lima, Lourenco, and Risau-Gusman}}]{Kinouchi_thesaurus:2001}
\bibinfo{author}{\bibfnamefont{O.}~\bibnamefont{Kinouchi}},
  \bibinfo{author}{\bibfnamefont{A.~S.} \bibnamefont{Martinez}},
  \bibinfo{author}{\bibfnamefont{G.~F.} \bibnamefont{Lima}},
  \bibinfo{author}{\bibfnamefont{G.~M.} \bibnamefont{Lourenco}},
  \bibnamefont{and}
  \bibinfo{author}{\bibfnamefont{S.}~\bibnamefont{Risau-Gusman}},
  \bibinfo{journal}{Physica A} \textbf{\bibinfo{volume}{315}},
  \bibinfo{pages}{665} (\bibinfo{year}{2002}).

\bibitem[{\citenamefont{Yang}(2005)}]{Yang:2005}
\bibinfo{author}{\bibfnamefont{S.~J.} \bibnamefont{Yang}},
  \bibinfo{journal}{Phys. Rev. E} \textbf{\bibinfo{volume}{71}},
  \bibinfo{pages}{016107} (\bibinfo{year}{2005}).

\bibitem[{\citenamefont{da~F.~Costa}(2006)}]{Costa_know:2006}
\bibinfo{author}{\bibfnamefont{L.}~\bibnamefont{da~F.~Costa}},
  \bibinfo{journal}{Phys. Rev. E} \textbf{\bibinfo{volume}{74}},
  \bibinfo{pages}{026103} (\bibinfo{year}{2006}).

\bibitem[{\citenamefont{Herrero}(2003)}]{Herrero_self:2003}
\bibinfo{author}{\bibfnamefont{C.~P.} \bibnamefont{Herrero}},
  \bibinfo{journal}{Phys. Rev. E} \textbf{\bibinfo{volume}{68}},
  \bibinfo{pages}{026106} (\bibinfo{year}{2003}).

\bibitem[{\citenamefont{Herrero and Saboya}(2005)}]{Herrero_self:2005}
\bibinfo{author}{\bibfnamefont{C.~P.} \bibnamefont{Herrero}} \bibnamefont{and}
  \bibinfo{author}{\bibfnamefont{M.}~\bibnamefont{Saboya}},
  \bibinfo{journal}{Phys. Rev. E} \textbf{\bibinfo{volume}{71}},
  \bibinfo{pages}{016103} (\bibinfo{year}{2005}).

\bibitem[{\citenamefont{da~F.~Costa}(2008)}]{Costa_diverse:2008}
\bibinfo{author}{\bibfnamefont{L.}~\bibnamefont{da~F.~Costa}}
  (\bibinfo{year}{2008}), \bibinfo{note}{arXiv:0801.0380}.

\bibitem[{\citenamefont{da~F.~Costa}(2007{\natexlab{a}})}]{Costa_comp:2007}
\bibinfo{author}{\bibfnamefont{L.}~\bibnamefont{da~F.~Costa}}
  (\bibinfo{year}{2007}{\natexlab{a}}), \bibinfo{note}{arXiv:0711.2736}.

\bibitem[{\citenamefont{da~F.~Costa
  et~al.}(2007{\natexlab{b}})\citenamefont{da~F.~Costa, Rodrigues, Travieso,
  and Boas}}]{Costa_surv:2007}
\bibinfo{author}{\bibfnamefont{L.}~\bibnamefont{da~F.~Costa}},
  \bibinfo{author}{\bibfnamefont{F.~A.} \bibnamefont{Rodrigues}},
  \bibinfo{author}{\bibfnamefont{G.}~\bibnamefont{Travieso}}, \bibnamefont{and}
  \bibinfo{author}{\bibfnamefont{P.~R.~V.} \bibnamefont{Boas}},
  \bibinfo{journal}{Advs. in Phys.} \textbf{\bibinfo{volume}{56}},
  \bibinfo{pages}{167} (\bibinfo{year}{2007}{\natexlab{b}}).

\bibitem[{\citenamefont{Albert and Barab\'asi}(2002)}]{Albert_Barab:2002}
\bibinfo{author}{\bibfnamefont{R.}~\bibnamefont{Albert}} \bibnamefont{and}
  \bibinfo{author}{\bibfnamefont{A.~L.} \bibnamefont{Barab\'asi}},
  \bibinfo{journal}{Rev. Mod. Phys.} \textbf{\bibinfo{volume}{74}},
  \bibinfo{pages}{47} (\bibinfo{year}{2002}).

\bibitem[{\citenamefont{Dorogovtsev and Mendes}(2002)}]{Dorogov_Mendes:2002}
\bibinfo{author}{\bibfnamefont{S.~N.} \bibnamefont{Dorogovtsev}}
  \bibnamefont{and} \bibinfo{author}{\bibfnamefont{J.~F.~F.}
  \bibnamefont{Mendes}}, \bibinfo{journal}{Advs. in Phys.}
  \textbf{\bibinfo{volume}{51}}, \bibinfo{pages}{1079} (\bibinfo{year}{2002}).

\bibitem[{\citenamefont{Flory}(1941)}]{Flory}
\bibinfo{author}{\bibfnamefont{P.~J.} \bibnamefont{Flory}},
  \bibinfo{journal}{Journal of the American Chemical Society}
  \textbf{\bibinfo{volume}{63}}, \bibinfo{pages}{3083} (\bibinfo{year}{1941}).

\bibitem[{\citenamefont{da~F.~Costa}(2007{\natexlab{b}})}]{Costa_path:2007}
\bibinfo{author}{\bibfnamefont{L.}~\bibnamefont{da~F.~Costa}}
  (\bibinfo{year}{2007}{\natexlab{b}}), \bibinfo{note}{arXiv:0711.1271}.

\bibitem[{\citenamefont{da~F.~Costa}(2007{\natexlab{c}})}]{Costa_longest:2007}
\bibinfo{author}{\bibfnamefont{L.}~\bibnamefont{da~F.~Costa}}
  (\bibinfo{year}{2007}{\natexlab{c}}), \bibinfo{note}{arXiv:0712.0415}.

\bibitem[{\citenamefont{Latora and Baranger}(1999)}]{Latora_entropy}
\bibinfo{author}{\bibfnamefont{V.}~\bibnamefont{Latora}} \bibnamefont{and}
  \bibinfo{author}{\bibfnamefont{M.}~\bibnamefont{Baranger}},
  \bibinfo{journal}{Physical Review Letters} \textbf{\bibinfo{volume}{82}},
  \bibinfo{pages}{520} (\bibinfo{year}{1999}).

\bibitem[{\citenamefont{da~F.~Costa}(2004)}]{Costa_genperc}
\bibinfo{author}{\bibfnamefont{L.}~\bibnamefont{da~F.~Costa}},
  \bibinfo{journal}{Phys. Rev. E} \textbf{\bibinfo{volume}{70}},
  \bibinfo{pages}{056106} (\bibinfo{year}{2004}).

\bibitem[{\citenamefont{da~F.~Costa and Cesar}(2001)}]{Costa_book:2001}
\bibinfo{author}{\bibfnamefont{L.}~\bibnamefont{da~F.~Costa}} \bibnamefont{and}
  \bibinfo{author}{\bibfnamefont{R.~M.} \bibnamefont{Cesar}},
  \emph{\bibinfo{title}{Shape Analysis and Classification: {T}heory and
  Practice}} (\bibinfo{publisher}{CRC Press}, \bibinfo{year}{2001}).

\end{thebibliography}
\end{document}